\documentclass[11pt,twoside,a4paper]{article}
\usepackage[utf8]{inputenc}
\usepackage{latexsym,amsthm}
\usepackage{graphics,physics}
\usepackage{graphicx}
\usepackage{epsfig}
\usepackage{multirow}
\usepackage{amsmath,amssymb,bm}
\usepackage{hhline}
\usepackage{hyperref}
\usepackage{enumitem}
\usepackage{amsfonts}
\usepackage{xcolor}
\usepackage{soul}
\usepackage[font=small,labelfont=bf]{caption}
\usepackage[compat=1.1.0]{tikz-feynman}
\usepackage{subcaption}
\usepackage{hhline}
\usepackage{bm,ulem}
\usepackage{subcaption}
\usepackage{float}
\restylefloat{table}
\usepackage[margin=1in]{geometry}
\usepackage[compat=1.1.0]{tikz-feynman}
\usepackage{braket}
\usepackage{nccmath}
\usepackage{cite}
\usepackage{multirow}
\usepackage{makecell}
\def\plusheight{-\the\dimexpr\fontdimen22\textfont2\relax}
\usepackage{amsmath, amsthm}
\usepackage{subcaption}
\usepackage{multirow}
\usepackage{booktabs}
\usepackage{amsmath}
\usepackage{amsfonts}
\usepackage{amssymb}
\usepackage{graphicx}
\usepackage{indentfirst}
\usepackage{slashed}
\usepackage{mathrsfs}
\usepackage{feynmf}
\usepackage{comment}

\newcommand{\Disc}[0]{\operatorname{Disc}}
\newcommand{\Cut}[0]{\operatorname{Cut}}

\newcommand{\zbar}{\bar{z}}

\newcommand{\bea}{\begin{eqnarray}}
	\newcommand{\eea}{\end{eqnarray}}
\newcommand{\bean}{\begin{eqnarray*}}
	\newcommand{\eean}{\end{eqnarray*}}

\def\half{\frac{1}{2}}

\def\abs#1{\left| #1\right|}

\def\th{{\theta}}

\def\eps{\epsilon}

\def\cS{{\cal S}}

\def\Label#1{\label{#1}
	\smash{\hbox to0pt{\raise1ex\hbox{\tiny[#1]}\hss}}}

\def\beq{\begin{equation}}
	\def\eeq{\end{equation}}
\def\bsp#1\esp{\begin{split}#1\end{split}}

\newcommand{\cA}{\mathcal{A}}
\newcommand{\cH}{\mathcal{H}}

\renewcommand{\ln}{\log}

\title{\bf The Hopf Algebra Structure of the Two Loop Three Mass Non-Planar Feynman Diagram}

\author{\bf B. Ananthanarayan$^a$, \bf Abhijit B. Das$^a$, \bf Daniel Wyler$^b$} 

\date{%
	$^a$Centre for High Energy Physics,\\ Indian Institute of Science,\\ Bangalore-560012, Karnataka, India
	\\[\baselineskip]
	$^b$ Institute for Theoretical Physics \\
	University of Zürich \\
	Winterthurerstr. 190 \\
	8057 Zürich
	Switzerland }

\begin{document}
	\maketitle
	\begin{abstract}
		The method of using Hopf algebras for calculating Feynman integrals developed
		by Abreu et al. is applied to the two-loop non-planar on-shell diagram
		with massless propagators and three external mass scales. We show that the existence of the method of cut Feynman diagrams comprising of the coproduct,
		the first entry condition and integrability condition that was found to be true for the planar case also holds for
		the non-planar case; furthermore, the non-planar symbol alphabet is the
		same as for the planar case.  This is one of  the main results of this work, and they have
		been obtained by a systematic analysis of the relevant cuts, using the
		symbolic manipulation codes HypExp and PolyLogTools.  The obtained result for the symbol is cross-checked by an analysis of the known two-loop original Feynman integral result. In addition, we also reconstruct the full result  from the symbol. This is the other main result in this paper.
	\end{abstract}
	\section{Introduction}
	
	In a series of publications \cite{Abreu et al.(2014), Abreu et al.(2015), Abreu et al.(2017), Abreu et al.(2017)A}, Abreu et al. have developed the study of cut Feynman diagrams with the intention to exploit generalized unitarity, as encoded in the corresponding Hopf algebras, to calculate Feynman integrals in those cases where the final results and the corresponding cuts can be expressed by multiple polylogarithms. In fact, using the properties of the Hopf algebra, the 
	Feynman diagrams one seeks can be reconstructed from the cut Feynman integrals without the need to perform a tedious dispersion integral, although, in practice, there is still a considerable calculational effort required even in the evaluation of the cut diagrams. In passing, we note that in the case of, say, elliptic polylogarithms (e.g., the two-loop sunset), there is as yet no known (Hopf) algebraic structure that would simplify the dispersion integral.  In an important paper, Duhr \cite{Duhr:2012fh} has argued that for the large class of Feynman that result in multiple polylogarithms, the method of Hopf algebras can be applied.  It is not trivial that abstract mathematical objects like Hopf algebras 
	can be found in a real 
	physical system and it is of great interest to further study the extent	to which the new method applies in order to discover new calculational tools.
	The work described in this paper indeed intends to	contribute further to the method of using Hopf Algebras in the context of Feynman integrals.

	While at the one-loop level, the calculations are straightforward, the corresponding diagrams also exhibit important and useful mathematical properties such as the integrability condition and the first-entry condition made evident by the new formalism.  These are indeed related to the discovery that Feynman integrals obey Hopf Algebras, which plays an important role in the new method. At two-loop and higher orders, the level of complexity is much higher in terms of the number of integrations for a particular cut integral and the existence of divergences. The appearance of the delta function is what is particular to the cut method (Cutkosky condition).  The general advantage of
	the method is that one effectively computes lower-loop diagrams, and integrations of cut diagrams are simpler.
	
	In order to set up the formalism, the authors introduced and developed a methodology that is encoded in the following concepts: multiple polylogarithms, Hopf algebra, symbol alphabet, integrability condition, first-entry condition, and coproduct. In the next section, each of the relevant terms is explained. In ref.\cite{Abreu et al.(2014)}, besides several one-loop 
	examples, the two-loop (planar) ladder diagram is considered.
	This diagram, which has been studied in the past and has received a lot of attention, is a very useful example for illustrative purposes.  It is finite and can be expressed in closed form in 4 dimensions \cite{Usyukina:1992jd}.
	The calculation using the new method involves several
	steps: First of all, the total cut diagram for any one channel (of the external momenta), which is a sum of all the possible cuts (the cut propagator momenta add up to the channel momentum), needs to be evaluated for that channel. Then, using this first result,  the maximal iteration of the coproduct or the symbol is evaluated using the first-entry condition and the integrability condition. Finally, the original Feynman integral is reconstructed by finding the class of Feynman integrals, which can give rise to the symbol. This is done by explicitly singling out the possible expressions which can give rise to each of the terms in the symbol and finally adding them up to get the class.

\begin{figure}[]
		\begin{subfigure}[]{0.48\linewidth}
			\centering
			\includegraphics[keepaspectratio=true, height=6cm]{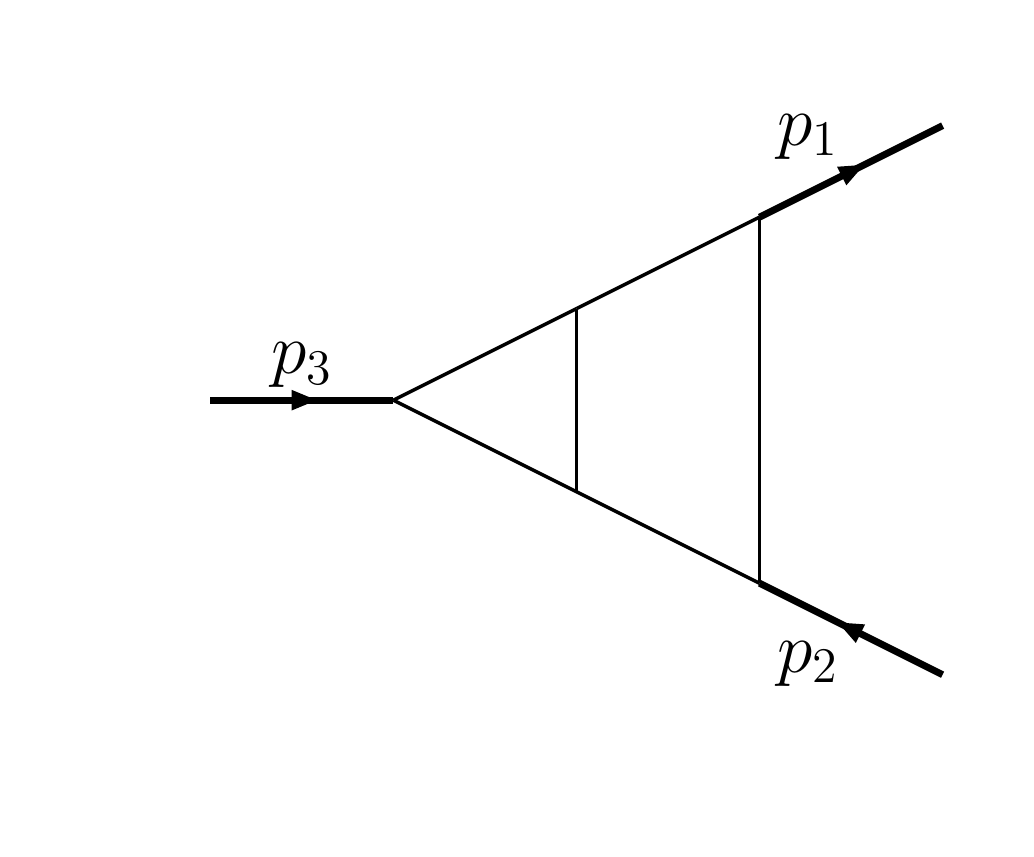}
			\caption{$\textrm{Ladder diagram}$}\label{ladder}
		\end{subfigure}
		\begin{subfigure}[]{0.48\linewidth}
			\centering
			\includegraphics[keepaspectratio=true, height=6cm]{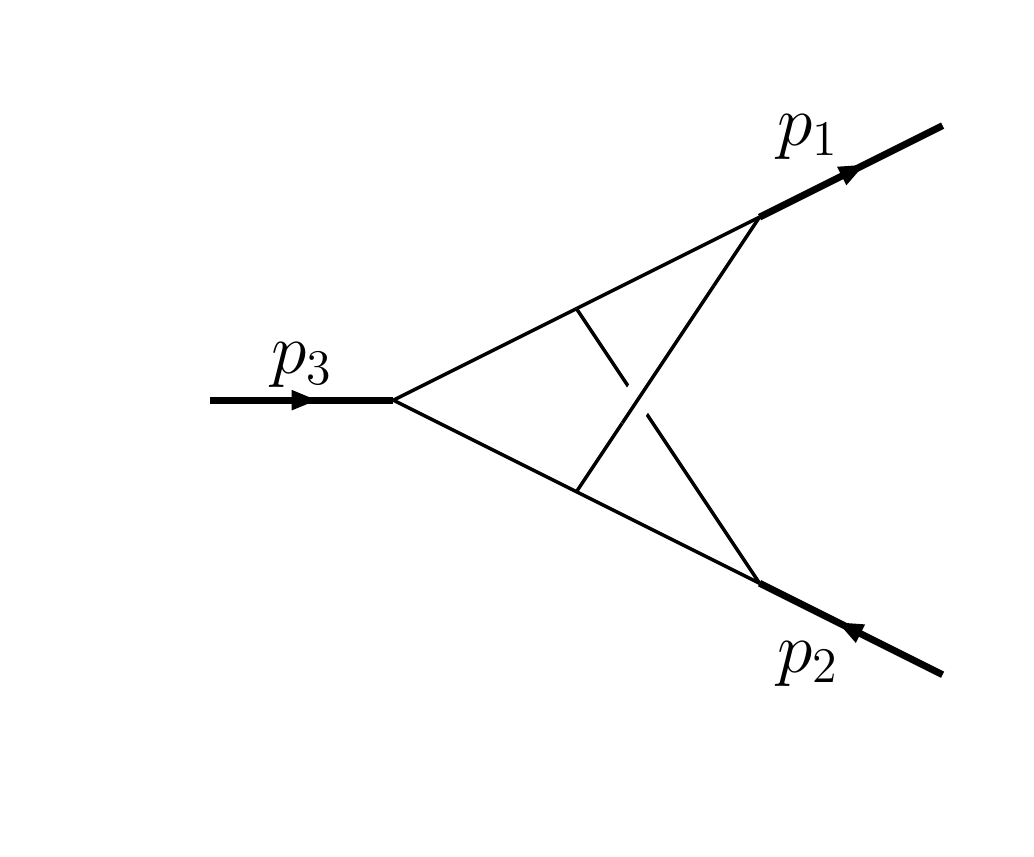}
			\caption{$\textrm{Non-planar diagram}$}\label{nladder}
		\end{subfigure} 
		\caption{The planar and non-planar two-loop diagrams discussed in this paper}
	\end{figure}

	In ref.\cite{Abreu et al.(2014)} it is pointed out that no non-planar diagram has been evaluated by themselves using the present Hopf Algebra based method. A non-planar `ladder' diagram has been calculated by Ussyukina and
	Davydychev \cite{Ussyukina Davydychev(1993)}, in addition to the planar topology.  The result is known in closed form, which is finite in 4 dimensions. In view of this considerations, we fell it is a natural choice for study using the Hopf algebra method. 
	We also note that, more recently, non-planar 
	examples with a related technology involving the symbol alphabet have been studied in \cite{chicherin,Zeng,mitev,dixon}. However, in addition to the symbol, 
	these authors also involve Mellin-Barnes technology and differential equations to compute the non-planar integrals. We intend to rely only on the symbols in order to expand their range of use. The diagrams are given by Fig.(\ref{ladder}) and Fig.(\ref{nladder}).
	
	The availability of this result offers the possibility of checking whether or not the
	Hopf algebra results of the planar case also carry over to the non-planar case.
	The objective of this paper is to answer this question; it is found to be in the affirmative. 
	%
	%	The conjecture used by Abreu et al.  applies to all Feynman integrals, which are multiple polylogarithms, whether planar 
	%	or non-planar. This work is a cross-check whether `the method' works for this non-planar case. However, it is not a proof for
	%	other non-planar cases. A justification proof for the above conjecture, particular to the Feynman integrals being multiple polylogarithms case, is given in \cite{Duhr:2012fh}. This proof applies to any Feynman integral being a multiple polylogarithm, whether it is planar, non-planar, or a multiloop integral.
	%	
	
	In order to tackle the above question, we have studied the cuts of this non-planar diagram in the $p_2^2$ channel. 
	There are two cuts possible for this channel, a two-propagator cut (with two propagators on-shell) and a three-propagator cut (with three propagators on-shell). For evaluating the integrals associated with the first cut and the  second cut, we have used the one-loop results of \cite{Abreu et al.(2017)A} and \cite{Duplancic n nizic} respectively. The net cut across the $p_2^2$ channel is the sum of these two cuts. The expression for this total cut is one of the significant result of this paper. (We have chosen the $p_2^2$ channel because its more simpler than the $p_3^2$ channel in terms of the number of cuts. The final result is independent of the choice of the channel.)
	Using this, we have
	reconstructed the original Feynman diagram. The reason behind choosing this diagram is the similarity in topology. It is a sufficiently simple extension of the planar ladder, but
	in practice, far more complicated.  
	
	The plan of this paper is the following: In Sec.2, we give a brief overview of the method given in Abreu et al.. Here first of all we give the definitions of multiple polylogarithms \cite{modular,tate}, their co-product \cite{tate,Galois} and the symbol \cite{Wilson,tate}.  After this, we show the equivalence of the coproduct with the symbol and then tell about the
	integrability condition
	\cite{Wilson,itin,Grass,zeta,polygons}. Then we give the relation between the discontinuity, cut, and the coproduct \cite{cutkosky,diagrammar,diagrammatica,Hopf}, and the definition of the first-entry condition \cite{landau,straps} and finally, the reconstruction of the symbol and the original Feynman integral. 
	In Sec. 3, we give the diagram and the kinematics.
	In Sec. 4, we evaluate the first kind of cut for the $p_2^2$ channel.  We follow the same conventions of Abreu et al. for parametrizing the momentum variables. The symbol alphabet is the same as discussed in \cite{Abreu et al.(2014)} for the three mass triangle and the two-loop ladder diagram owing to the fact that the non-plnar integral falls into the same family of integrals considered in \cite{Ussyukina Davydychev(1993)}.  In Sec. 5, we evaluate the second kind of cut for the $p_2^2$ channel.  In Sec. 6, we do the sum of the results of  the two cuts giving the total cut result  across the $p_2^2$ channel to be finite. Thus we have shown that since the full result given in \cite{Ussyukina Davydychev(1993)} is divergenceless, so is the cut result.   In Sec. 7, we reconstruct the symbol of the original Feynman diagram using the first-entry condition and the integrability condition. From the known result of \cite{Ussyukina Davydychev(1993)} we compute the symbol in order to verify the obtained result. 
	The reconstruction of the symbol verifies the Duhr conjecture \cite{Duhr:2012fh} for the two loop non planar triangle diagram and is one of the main results of this paper. In other words, for the non-planar example, the Hopf Algebra structure is preserved.
	In Sec. 8, we reconstruct the full function (final result of the diagram).  The reconstruction of the full function involves using a systematic search for all the harmonic polylogarithms (HPL's) \cite{HPL,BrownSVHPLs} of relevant weights and imposing the constraints coming from the Cutkosky rules.  In Sec.9,
	we give our conclusions and a discussion. The Appendix describes our technical calculations involving the cancellation of the divergences.
	
	We note here that at the time the work of Abreu et al. was performed PolyLogTools \cite{PolylogTools} package was not publicly available.  The availability of the code has rendered this evaluation possible by us.  In addition, we have used the HypExp \cite{Hypexp} package. 
	It is of note that a new work extending the diagrammatic coaction already established at one loop level \cite{Abreu et al.(2017)A} has now been extended to some two loop examples \cite{diagrammatictwoloop} yet. Applications to non-planar diagrams is yet to be explored.

	\section{Brief overview of the method of cut Feynman diagrams}
	
	Here we give a brief discussion of the method of cut Feynman diagrams given in \cite{Abreu et al.(2014)} worked out for the planar examples. It is a rephrasing of the known information that is given here for completeness. The interested reader is urged to go through the original references in order to have a complete picture. 
	\subsection{Introduction}
	A large class of Feynman integrals can be expressed in terms of transcendental functions called multiple polylogarithms, which are defined by certain iterated integrals and include classical polylogarithms as a special case.
	Multiple polylogarithms form a Hopf algebra which involves a coproduct $\Delta$. 
	The discontinuity across a channel of the Feynman integral is related to the coproduct. 
	The coproduct, in turn, is related to the symbol alphabet of the Feynman integral function, which encodes the information about this function. 
	Thus doing the simpler task of analytically evaluating the discontinuity of the integral can lead us to determine the original Feynman integral. 
	\subsection{Multiple polylogarithms}
	Multiple polylogarithms are defined by the iterated integral \cite{tate,modular}
	\beq
	G(a_1,\ldots,a_n;z)=\,\int_0^z\,{d t\over t-a_1}\,G(a_2,\ldots,a_n;t)\,,
	\eeq
	with $a_i, z\in \mathbb{C}$.  
	The number $n$ of integrations is called the \textit{weight} of the multiple polylogarithm. 
	We denote by $\overline\cH$ the $\mathbb{Q}$-vector space spanned by all multiple polylogarithms, which can be turned into an algebra as well.  
	The algebra of multiple polylogarithms is \textit{graded} by the weight,
	\beq
	\overline\cH = \bigoplus_{n=0}^\infty\overline\cH_n{\rm~~with~~} \overline\cH_{n_1}\cdot\overline\cH_{n_2}\subset\overline\cH_{n_1+n_2}\,,
	\eeq
	where $\overline\cH_n$ is the $\mathbb{Q}$-vector space spanned by all multiple polylogarithms of weight $n$, and we define $\overline\cH_0=\mathbb{Q}$.
	
	%%%%%%%%%%%%%%%%%%%%%%%%%%%%%%%%%%%%%%%%%%%%%%%%%%%%%%%%%%%%%%%%%%%%%%%%%%%%%%%%%%
	
	\subsection{The Coproduct}
	Multiple polylogarithms can be endowed with surprising algebraic structures. 
	They obey the quotient space $\cH = \overline\cH/(\pi\,\overline\cH)$ (the algebra $\overline\cH$ modulo $\pi$), where $\cH$ is a Hopf algebra \cite{Galois,tate}. 
	$\cH$ can be equipped with a \textit{coproduct} $\Delta:\cH\to\cH\otimes\cH$, which is coassociative,
	\beq\label{idco}
	(\textrm{id}\otimes\Delta)\,\Delta = (\Delta\otimes\textrm{id})\,\Delta\,,
	\eeq
	
	with the multiplication rule,
	\beq
	\Delta(a\cdot b) = \Delta(a)\cdot\Delta(b)\,,
	\eeq
	
	and the weight rule,
	\beq
	\cH_n  \stackrel{\Delta}{\longrightarrow} \bigoplus_{k=0}^n\cH_k\otimes\cH_{n-k}\,.
	\eeq
	The co-product of the multiple polylogarithms is defined by \cite{Galois}
	\begin{align}\label{eq:coproduct}
		\Delta(G(a_0;a_1,\ldots,a_n;a_{n+1})) &\,= \sum_{0=i_1<i_2<\ldots<i_{k}<i_{k+1}=n}\!\!\! G(a_0;a_{i_1},\ldots,a_{i_k};a_{n+1})\\
		\nonumber&\qquad \otimes\Bigg[\prod_{p=0}^kG(a_{i_p};a_{i_p+1},\ldots,a_{i_{p+1}-1};a_{i_{p+1}})\Bigg]\,.
	\end{align}
	Using this, the coproduct of the ordinary logarithm and the classical polylogarithms are 
	\beq
	\Delta(\log z) = 1\otimes\log z+\log z\otimes 1 {\rm~~and~~}
	\Delta(\textrm{Li}_n(z)) = 1\otimes\textrm{Li}_n(z) + \sum_{k=0}^{n-1}\textrm{Li}_{n-k}(z)\otimes\frac{\log^kz}{k!}\,.
	\eeq
	Also
	\beq\bsp
	\Delta(\ln x\ln y) &\,= \Delta(\ln x)\,\Delta(\ln y) = [1\otimes \ln x+\ln x\otimes 1]\,[1\otimes \ln y+\ln y\otimes 1]\\
	&\, = 1\otimes(\ln x\,\ln y) + \ln x\otimes \ln y + \ln y\otimes \ln x + (\ln x\,\ln y)\otimes 1\,.
	\esp\eeq
	
	If $(n_1,\ldots,n_k)$ is a partition of $n$, we define
	\beq
	\Delta_{n_1,\ldots,n_k} : \cH_{n}\to\cH_{n_1}\otimes\ldots\otimes\cH_{n_k}\,.
	\eeq
	
	Using eq.(6) we can then write
	\beq\label{sumco}
	\Delta= \sum_{p+q=n}\Delta_{p,q}\,,
	\eeq
	and it satisfies the recursion
	\beq\label{eq:iterated_Delta}
	\Delta_{n_1,\ldots,n_k} = (\Delta_{n_1,\ldots,n_{k-1}}\otimes \textrm{id})\Delta_{n,n_k}\,,\qquad n=n_1+\ldots+n_{k-1}\,.
	\eeq
	%%%%%%%%%%%%%%%%%%%%%%%%%%%%%%%%%%%%%%%%%%%%%%%%%%%%%%%%%%%%%%%%%%%%%%%%%%%%%%%%%%
	
	\subsection{The Symbol}
	The symbol of a transcendental function $F_w(x_1,\ldots,x_n)$ of weight $w$ in the variables $x_1,\ldots,x_n$ is defined recursively by \cite{Wilson}
	\beq
	\cS(F_w) = \sum_i\cS(F_{i,w-1})\otimes R_i\,.
	\eeq
	where $F_{i,w-1}$ are transcendental functions of weight $w-1$ and the $R_i$ are rational functions in the variables $x_1,\ldots,x_n$, with the condition that the total differential of $F_w$ can be written in the form
	\beq\label{eq:diff_eq_generic}
	dF_w = \sum_iF_{i,w-1}\,d\ln R_i\,,
	\eeq
	
	Multiple polylogarithms satisfy a differential equation of the type~eq.\eqref{eq:diff_eq_generic} \cite{tate},
	\beq\label{eq:dif_eq_polylog}
	dG(a_0;a_1,\ldots,a_n;a_{n+1}) = \sum_{i=1}^nG(a_0;a_1,\ldots,\hat{a}_i,\ldots,a_n;a_{n+1})\,d\ln\left({a_{i+1}-a_i\over a_{i-1}-a_i}\right)\,,
	\eeq
	where the hat indicates that the corresponding element is omitted.
	
	\subsection{Equivalence of the co-product with the symbol and the integrability Condition}
	The maximal iteration of the coproduct, corresponding to the partition $(1,\ldots,1)$, is equivalent with the symbol of a transcendental function $F$ \cite{itin,Wilson,Grass,zeta,polygons}
	\beq
	\cS(F) \equiv \Delta_{1,\ldots,1}(F)\in\cH_1\otimes\ldots\otimes\cH_1\,.
	\eeq
	
	Not every element in $\cH_1\otimes\ldots\otimes\cH_1$corresponds to the symbol of a function in $\cH$, it should satisfy the
	\textbf{integrability condition :} If we take an element
	\beq
	s = \sum_{i_1,\ldots,i_n}c_{i_1,\ldots,i_n}\,\log x_{i_1}\otimes\ldots\otimes \log x_{i_n} \in\cH_1\otimes\ldots\otimes\cH_1\,,
	\eeq
	then there is a function $F\in\cH_n$ such that $\cS(F) = s$ if and only if $s$ satisfies the \textit{integrability condition}
	\beq\label{eq:integrability}
	\sum_{i_1,\ldots,i_n}c_{i_1,\ldots,i_n}\,d\log x_{i_k}\wedge d\log x_{i_{k+1}}\,\log x_{i_1}\otimes\ldots\otimes\log x_{k-1}\otimes\log x_{k+2}\otimes\ldots\otimes \log x_{i_n} = 0\,,
	\eeq
	where $\wedge$ denotes the usual wedge product on differential forms.
	
	\subsection{Disc and Cut}
	The operator $\Disc_s F$ gives the direct value of the discontinuity of $F$ as the variable $s$ crosses the real axis.
	\bea
	\Disc_s \left[F(s\pm i 0)\right]= \lim_{\varepsilon \to 0}\left[F(s\pm i \varepsilon)-F(s \mp i \varepsilon)\right],
	\label{eq:def-disc}
	\eea
	
	The operator $\Cut_s$ gives the sum of \textit{cut} Feynman integrals, in which some propagators in the integrand of $F$ are replaced by Dirac delta functions. For a massless sacalar theory, the rules for cut may be depicted as :
	\begin{align}
		\raisebox{-0.1mm}{\includegraphics[keepaspectratio=true, height=0.2cm]{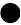}}=i
		\qquad \qquad \qquad
		\raisebox{-0.1mm}{\includegraphics[keepaspectratio=true, height=0.2cm]{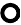}}=-i
	\end{align}
	\begin{align}
		\raisebox{-0.1mm}{\includegraphics[keepaspectratio=true, height=0.55cm]{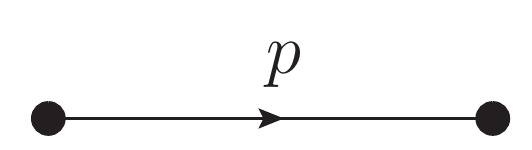}}=\frac{i}{p^2+i\varepsilon}
		\qquad \qquad
		\raisebox{-0.1mm}{\includegraphics[keepaspectratio=true, height=0.55cm]{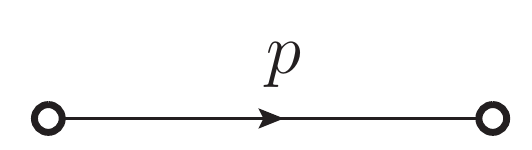}}=\frac{-i}{p^2-i\varepsilon}
	\end{align}
	\begin{align}\label{cutdefdiag}
		\raisebox{-6.7mm}{\includegraphics[keepaspectratio=true, height=1.6cm]{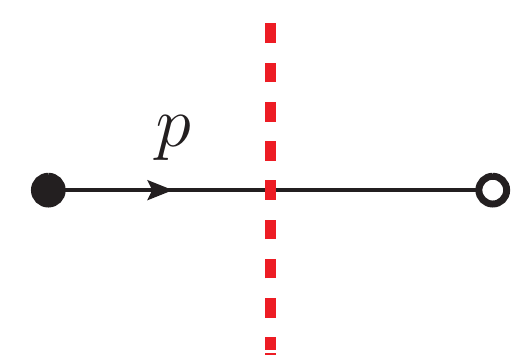}}
		= 2\pi\, \delta\left(p^2\right)\theta\left(p_0\right)
	\end{align}
	%%%%%%%%%%%%%%%%%%%%%%%%%%%%%%%%%%%%%%%%%%%%%%%%%%%%%%%%%%%
	
	\subsection{Relations between Disc, Cut and coproduct}
	Cutkosky's rule is given by \cite{cutkosky,diagrammar,diagrammatica} :
	\bea
	\Disc_s F =-\Cut_s F.
	\label{eq:oldcutting}
	\eea
	
	The Disc and Coproduct are related by \cite{Hopf}
	\bea
	\Disc f_n \cong \mu\left[(\Disc \otimes \textrm{id})\,\Delta_{1,n-1} f_n\right]\,
	\label{eq:disc-with-mu}
	\eea
	where $\mu:\overline\cH\otimes\overline\cH\to\overline\cH$ denotes the multiplication in $\overline\cH$, i.e.\ we simply multiply the two factors in the coproduct, and $\cong$ denotes equivalence modulo $\pi^2$. 
	Thus from eq.(\ref{eq:oldcutting}) and eq.(\ref{eq:disc-with-mu}) we have
	\bea
	\Cut f_n \cong -\mu\left[(\Disc \otimes \textrm{id})\,\Delta_{1,n-1} f_n\right]\,
	\eea
	
	%%%%%%%%%%%%%%%%%%%%%%%%%%%%%%%%%%%%%%%%
	\subsection{The First-entry Condition}
	For massless propagators, the branch points of the integral, seen as a function of the invariants $s_{ij} = (p_i + p_j)^2$ are the points where one of the invariants is zero or infinite \cite{landau}. \\[\baselineskip] Using eq.(\ref{eq:disc-with-mu}) this implies the so-called \textbf{first entry condition}, i.e., the statement that the first entries of the symbol of a Feynman integral with massless propagators can only be logarithms of Mandelstam invariants \cite{straps} to give a non-zero value to the discontinuity of the original function.  \\[\baselineskip]
	Thus 
	\bea
	\Delta_{1,n-1} f_n = \log s_{ij} \otimes f_{n-1}
	\eea
	
	which implies
	\bea
	\Cut f_n \cong -\mu\left[(\Disc \otimes \textrm{id})\,(\log s_{ij} \otimes f_{n-1})\right]\,
	\eea
	
	and hence
	\bea
	\Cut f_n \cong  -\mu \left[\Disc (\log s_{ij}) \otimes f_{n-1}\right] = \Disc (\log s_{ij}) f_{n-1}
	\eea
	Thus the cut is proportional to the second entry of $\Delta_{1,n-1}$.
	%%%%%%%%%%%%%%%%%%%%%%%%%%%%%%%%%%%%%%%%%%%%%%%%%%%%%%%%%%%%%%%%	
	
	\subsection{Reconstructing the Symbol from the Cut}
	The Symbol can be obtained from the Cut using the following three conditions:  
	\begin{itemize}
		\item Cut is proportional to second entry of $\Delta_{1,n-1}$ 
		\item The First-entry condition 
		\item The Integrabilty condition  
	\end{itemize}
	First of all, we take the obtained result for the cut as the second-entry of $\Delta_{1,n-1}$ and the respective channel (the channel which has been cut) as its first-entry. Then using eq.(\ref{idco})-eq.(\ref{eq:iterated_Delta}), we can find out a possible first guess for the maximal iteration of the co-product $\Delta_{1,...,1}$ which is equal to the symbol. After this, we need to apply the integrability condition and the first entry condition repeatedly until both these conditions are satisfied, and finally, we get the symbol.
	%%%%%%%%%%%%%%%%%%%%%%%%%%%%%%%%%%%%%%%%%%%%%%%%%%%%%%%%%%%%%%%%%%%%%%%%%%%
	
	\subsection{Reconstructing the Feynman Integral from the symbol}
	After getting the symbol, we can guess 
	%not guess? determine?
	which kind of function has given rise to each term in the obtained symbol by application of the coproduct $\Delta_{1,...,1}$ on it. After getting all such functions, we add them up to get the family of functions 
	%the sum is a family?
	that give the obtained complete symbol. After this, we can use eq.(\ref{eq:oldcutting}) 
	%correct?
	to get the actual function out of the family of functions. 
	\begin{figure}[]
		\centering
		\includegraphics[keepaspectratio=true, height=10cm]{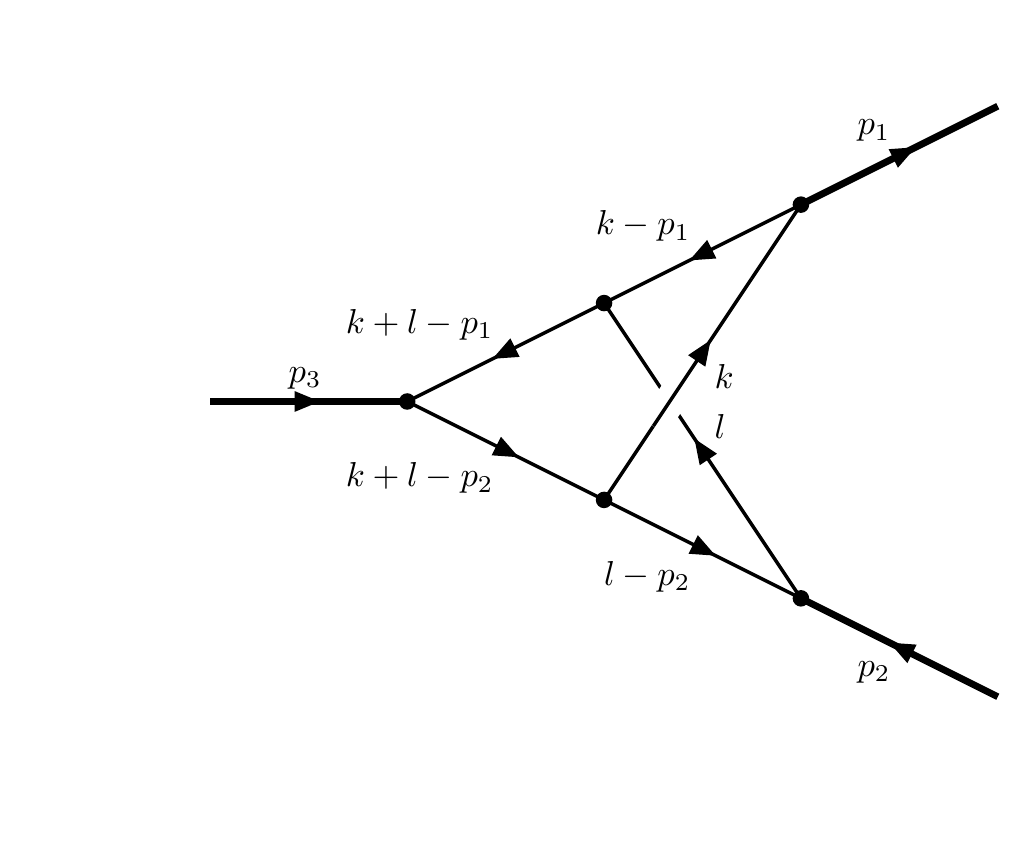}
		\caption{$C(p_1^2,p_2^2,p_3^3)$}\label{fig1}
	\end{figure}
	%%%%%%%%%%%%%%%%%%%%%%%%%%%%%%%%%%%%%%%%%%%%%%%%%%%%%%%%%%%%%%%%%%%%%%%%%%
	
	\section{The Two-loop non-planar Diagram}
	As we have pointed out, the above method has been checked only for planar topologies. Here we take a topologically related two-loop three mass non-planar diagram with massless internal propagators. It is illustrated in Fig.(\ref{fig1}) and the corresponding Feynman integral is given by
	\beq
	\label{nonplanar}
	C(p_1^2,p_2^2,p_3^3)=e^{2\gamma_E \eps}
	\int \frac{d^{D}l}{\pi^{D/2}}\frac{d^{D}k}{\pi^{D/2}}\, \frac{1 }{(l^2) \,\left((l-p_2)^2\right)(k+l-p_1)^2 (k+l-p_2)^2(k-p_1)^2(k)^2}\,,
	\eeq
	\\
	which can be evaluated using direct integration and is given by \cite{Ussyukina Davydychev(1993)}
	$$C(p_1^2,p_2^2,p_3^3)=$$
	\begin{equation}\label{orgfeyn}
		\left(\frac{\rm{i}\pi}{p_3^2}\right)^2\left(\frac{1}{\lambda}\left[2Li_2(-\rho x)-2Li_2(-\rho y)+\log(\rho x)\log(\rho y)+\log\left(\frac{y}{x}\right)\log\left(\frac{1+\rho y}{1+\rho x}\right)+\frac{\pi^2}{3}\right]\right)^2
	\end{equation}
	
	with
	\beq
	\lambda=\sqrt{(1-x-y)^2-4xy},\hspace{0.2cm}\rho(x,y)=\frac{2}{1-x-y+\lambda},\hspace{0.2cm} x=\frac{p_1^2}{p_3^2},\hspace{0.2cm} y=\frac{p_2^2}{p_3^2}
	\eeq
	It is worth emphasizing that the result can be expressed in terms of the same variables $z$ and $\zbar$ in \cite{Abreu et al.(2014)} which is used to express the results of the topologically related one-loop triangle and the two-loop ladder diagram. Now we will try to prove the above method for the $p_2^2$ channel cut. In a channel cut, the momenta of internal lines which are cut add up to the momentum of the channel. So using this, we have two types of cuts possible for this channel. The first cut, which we are going to consider, is a two-propagator cut and the second one is a three-propagator cut. \\
	\indent In contrast to the original Feynman integral being finite in $D=4$ dimensions, it turns out that the individual cut diagrams are divergent. Hence, we need to apply the dimensional regularization $D=4-2\eps$ in order to separate out the divergences in form of $\eps^{-1}$ terms. Also, in similarity with the two-loop ladder discussed in \cite{Abreu et al.(2014)}, the divergent terms from the individual cut diagrams cancel, giving the total cut finite, which is a required criterion as our original Feynman diagram is finite. \\ 
	\indent Also unlike in \cite{Abreu et al.(2014)}, here we have not considered a close examination of the kinematic regions explicitly where the kinematic invariants are of different sign for different regions. This is because the close inspection of these regions becomes essential only while considering the double unitary cuts and beyond, whereas, in our paper, we focus only on single unitary cuts for a diagram.
	\begin{figure}[]
		\begin{subfigure}[]{0.48\linewidth}
			\centering
			\includegraphics[keepaspectratio=true, height=7cm]{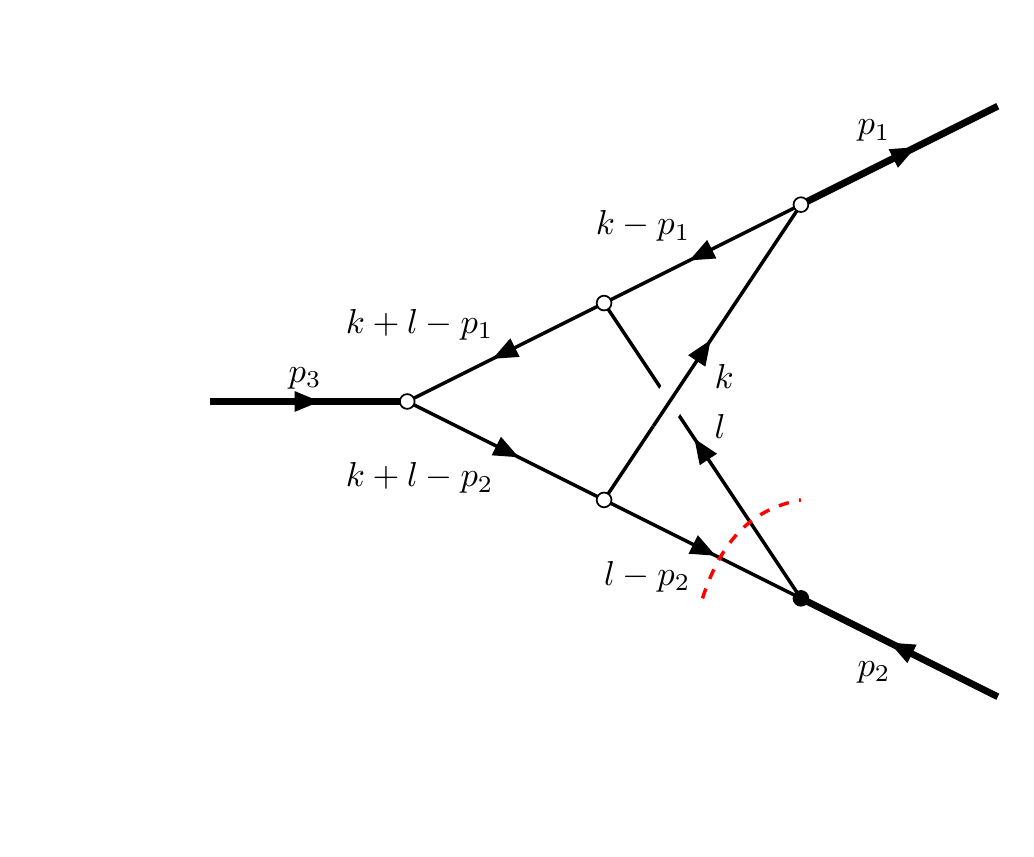}
			\caption{$\textrm{Cut}\hspace{0.05cm}\left[(l-p_2)^2,l^2\right]$}\label{fcd}
		\end{subfigure}
		\begin{subfigure}[]{0.48\linewidth}
			\centering
			\includegraphics[keepaspectratio=true, height=7cm]{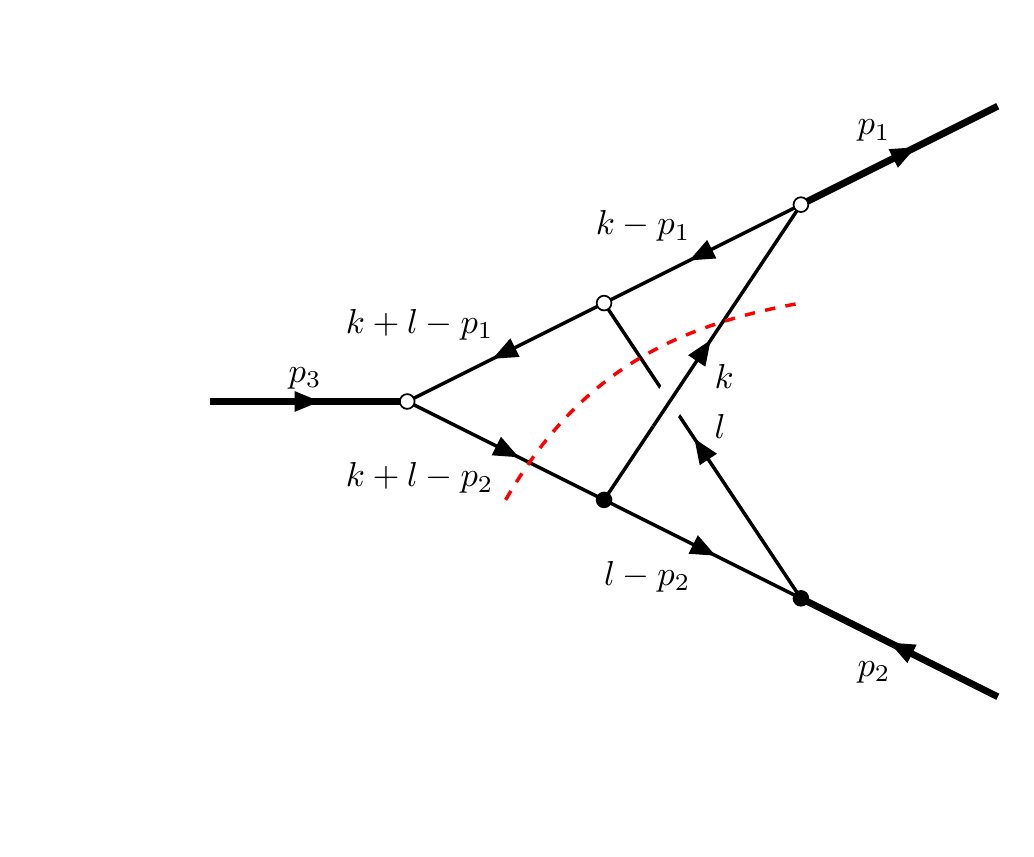}
			\caption{$\textrm{Cut}\hspace{0.05cm}\left[(k+l-p_2)^2,l^2,k^2\right]$}\label{scd}
		\end{subfigure} 
	\end{figure}
	
	%%%%%%%%%%%%%%%%%%%%%%%%%%%%%%%%%%%%%%%%%%%%%%%%%%%%%%%%%%%%%%%%%%%%%%%%%%%%%%%%%%
	\section{The First Cut}
	The Feynman integral for the first cut we are going to consider is given by
	\beq
	\label{nonplanarfirstcut}
	\textrm{Cut}\hspace{0.05cm}\left[(l-p_2)^2,l^2\right]=- (2\pi)^2\,e^{2\gamma_E \eps}
	\int \frac{d^{D}l}{\pi^{D/2}}\frac{d^{D}k}{\pi^{D/2}}\, \frac{ \delta^+(l^2) \,\delta^+\left((l-p_2)^2\right)}{(k+l-p_1)^2 (k+l-p_2)^2(k-p_1)^2(k)^2}\,,
	\eeq
	\\
	and the corresponding diagram is given in Fig.(\ref{fcd}). Here $\delta^+(l^2)=\delta(l^2)\theta(l_0)$ in accordance with eq.(\ref{cutdefdiag}). By close inspection of the diagram we can see that all the uncut propagators form a one loop two mass easy box and hence can be evaluated first independently. \\[\baselineskip]
	So the integral can be written as
	\beq\label{firstcut}
	\textrm{Cut}\hspace{0.05cm}\left[(l-p_2)^2,l^2\right]=\frac{e^{\gamma_E \epsilon}}{\pi^{2-\epsilon}}\int d^{4-2\epsilon}l(2\pi)^2\delta\left(l^2\right)
	\delta\left((l-p_2)^2\right)B^{2me}(s,t,p_1^2,p_3^2)\,.
	\eeq
	with $B^{2me}$ being the two mass easy box integral which is evaluated independently and is given by \cite{Abreu et al.(2017)A}
	\bea
	B^{2me}(s,t,p_1^2,p_3^2) &=&	{2c_{\Gamma} \over \eps^2(st-p_1^2p_3^2)}\left\{(-s)^{-\eps}+(-t)^{-\eps}-(-p_1^2)^{-\eps}-(-p_3^2)^{-\eps}\right. \nonumber \\ & & \left.+\sum_{j=0}^{3}(-1)^j\left({s+t-p_1^2-p_3^2  \over \alpha_j}\right)^{\eps}{}_2F_1\left(\eps;\eps;1+\eps;{st-p_1^2p_3^2 \over \alpha_j}\right)\right\} 
	\eea
	with
	\bea
	\alpha_0=(p_1^2-s)(p_1^2-t)\hspace{0.5cm}\alpha_1=(p_1^2-s)(s-p_3^2) \nonumber \\
	\alpha_2=(p_3^2-s)(p_3^2-t)\hspace{0.5cm}\alpha_3=(p_1^2-t)(t-p_3^2)
	\eea
	and
	\beq
	s=(p_3+l)^2,\hspace{0.2cm}t=(p_3-l+p_2)^2,\hspace{0.2cm}c_{\Gamma}={\Gamma(1+\eps)\Gamma^2(1-\eps)\over \Gamma(1-2\eps)}
	\eeq \\
	Note that here we have taken that expression for the result from literature which is complete for all orders in $\eps$, rather than taking the result where it is only upto $\eps^0$ order. This is because this result is actually more convenient for doing the subsequent calculations and to match the pre-factors of the two-different type of cuts which is required to do the cancellation of individual cut divergences. Now in similarity with the rules followed in \cite{Abreu et al.(2014)}, we will parametrize the momenta as
	\begin{align}\bsp
		&p_2 = \sqrt{p_2^2}(1,0,{\bf 0}_{D-2}), \qquad p_3 = \sqrt{p_3^2}(\alpha,\sqrt{\alpha^2-1},{\bf 0}_{D-2}),\\
		&l =  (l_0,|l| \cos \th,|l|\sin\th ~{\bf 1}_{D-2}),\label{eq:parametrization_nonplanar}
		\esp\end{align}
	where $\th \in [0,\pi]$ and $|l|>0$, and ${\bf 1}_{D-2}$ ranges over unit vectors in the dimensions transverse to $p_2$ and $p_3$. Note that this parametrization is possible because of the fact that three vectors can be constrained within a 3-dimensional co-ordinate space defined accordingly.
	Again using momentum conservation we have
	\bea
	\alpha = \frac{p_1^2-p_2^2-p_3^2}{2 \sqrt{p_2^2}\sqrt{p_3^2}}
	.
	\eea
	and using the parametrization above we rewrite the momentum integration as
	$$\frac{e^{\gamma_E \epsilon}}{\pi^{2-\epsilon}}\int d^{4-2\epsilon}l(2\pi)^2\delta\left(l^2\right)
	\delta\left((l-p_2)^2\right)=$$
	\beq
	\frac{4\pi e^{\gamma_E \epsilon}}{\Gamma(1-\epsilon)}\int dl_{0}\int d\abs{\bold{l}}^2
	\abs{\bold{l}}^{1-2\epsilon}\delta(l_{0}^2-\abs{\bold{l}}^2)\delta(p_2^2-2l_0\sqrt{p_2^2}+l^2)\int_{-1}^1 d\cos\theta (1-\cos^2\theta)^{-\epsilon}\,.
	\eeq
	\\[\baselineskip]
	The two delta functions allow us to trivially perform the $l_{0}$ and $\abs{\bold{l}}$ integrations. 
	\\[\baselineskip]
	The first integration enforces $\abs{\bold{l}}=l_{0}$ and the second one enforces $l_0={\sqrt{p_2^2}\over 2}$ everywhere in the integral with an additional factor of $2\sqrt{p_2^2}$ overall. 
	For the remaining integration, we will do the following  change of variables:
	\beq\label{cos}
	\cos \th = 2 x - 1, \qquad x \in [0,1],
	\eeq
	\beq\label{ui}
	u_i = \frac{p_i^2}{p_1^2}, \qquad i=2,3\,.
	\eeq
	\beq \label{zzbar}
	z = \half \left(1+u_2-u_3+\sqrt{\lambda}\right)\,,
	\qquad 
	\bar z = \half \left(1+u_2-u_3-\sqrt{\lambda}\right)\,,
	\eeq
	\beq \label{lambda}
	\lambda \equiv \lambda(1,u_2,u_3)\,,
	\qquad
	\lambda(a,b,c)=a^2+b^2+c^2-2ab-2ac-2bc\,,
	\eeq
	\\
	After the first two integrations are performed, with the above change of variables we have
	$$s=p_1^2[1-\bar z-x(z-\bar z)],\hspace{0.2cm}t=p_1^2[(1-z)+x(z-\bar z)],$$
	$$
	p_3^2=p_1^2[(1-z)(1-\bar z)],\hspace{0.2cm}p_2^2=p_1^2[z\bar z]
	$$
	and
	\begin{equation}
		s~ t-p_1^2p_3^2=(p_1^2)^2x(1-x)(z-\zbar)^2
	\end{equation}
	Finally eq.(\ref{firstcut}) becomes 
	$$
	\textrm{Cut}\hspace{0.05cm}\left[(l-p_2)^2,l^2\right]=(-1)^{\eps}\frac{4\pi c_{\Gamma} e^{\gamma_E \epsilon} }{\Gamma(1-\epsilon)}{(z\bar z)^{-\eps} \over (z-\bar z)^2}{(p_1^2)^{-2-2\eps} \over \eps^2}\int_{0}^1 dx\,x^{-1-\eps} (1-x)^{-1-\epsilon}\,
	$$
	$$
	\times \left\{ (1-\bar z-x(z-\bar z))^{-\eps}+(1-z+x(z-\bar z))^{-\eps}-(1)^{-\eps}-((1-z)(1-\bar z))^{-\eps}\right.
	$$
	$$
	+\left(z\bar z \over (\bar z+x(z-\bar z))(z-x(z-\bar z))\right)^{\eps}{}_2F_1\left(\eps;\eps;1+\eps;{x(1-x)(z-\bar z)^2 \over (\bar z+x(z-\bar z))(z-x(z-\bar z))}\right)
	$$
	$$
	-\left(z\bar z \over (\bar z+x(z-\bar z))(z(1-\bar z)-x(z-\bar z))\right)^{\eps}{}_2F_1\left(\eps;\eps;1+\eps;{x(1-x)(z-\bar z)^2 \over (\bar z+x(z-\bar z))(z(1-\bar z)-x(z-\bar z))}\right)
	$$
	$$
	-\left(z\bar z \over (\bar z(1-z)+x(z-\bar z))(z-x(z-\bar z))\right)^{\eps}{}_2F_1\left(\eps;\eps;1+\eps;{x(1-x)(z-\bar z)^2 \over (\bar z(1-z)+x(z-\bar z))(z-x(z-\bar z))}\right)
	$$
	$$
	+\left(z\bar z \over (\bar z(1-z)+x(z-\bar z))(z(1-\bar z)-x(z-\bar z))\right)^{\eps}
	$$
	\beq
	\left. \times {}_2F_1\left(\eps;\eps;1+\eps;{x(1-x)(z-\bar z)^2 \over (\bar z(1-z)+x(z-\bar z))(z(1-\bar z)-x(z-\bar z))}\right) \right\}
	\eeq
	\\
	So it turns out that we are able to express the cut totally in terms of the momentum invariant $p_1^2$ and the defined variables $z$ and $\zbar$ completely as expected. Now the first four integrations can be done by first expanding them in orders of epsilon and then integrating using PolyLogTools \cite{PolylogTools}. The last four integrations can be done by first expanding the Hypergeometric ${}_2F_1$ functions using HypExp \cite{Hypexp} and then integrating using PolyLogTools. 
	The results are as follows:
	\begin{align}
		\textrm{Cut}\hspace{0.05cm}\left[(l-p_2)^2,l^2\right]=(-1)^{\eps}\frac{4\pi c_{\Gamma} e^{\gamma_E \epsilon} }{\Gamma(1-\epsilon)}{(z\bar z)^{-\eps} \over (z-\bar z)^2}(p_1^2)^{-2-2\eps}\sum_{k=-1}^{\infty}\eps^k[f^{(k)}(z,\zbar)] 
	\end{align}
	with
	\begin{align}\label{firstcutdiv}
		&f^{(-1)}(z,\zbar)=-G\left(0,\frac{z-1}{z-\zbar},1\right)-G\left(0,\frac{\zbar-1}{\zbar-z},1\right)+G\left(1,\frac{z-1}{z-\zbar},1\right)+G\left(1,\frac{\zbar-1}{\zbar-z},1\right) \end{align}
	
	$$
	f^{(0)}(z,\zbar)=
	$$
	$$
	G(1,z) \left(G\left(0,\frac{z-1}{z-\zbar},1\right)-G\left(0,\frac{z}{z-\zbar},1\right)+G\left(0,\frac{z-z \zbar}{z-\zbar},1\right)-G\left(1,\frac{z-1}{z-\zbar},1\right)+\right. 
	$$
	$$
	\left.G\left(1,\frac{z}{z-\zbar},1\right)-G\left(1,\frac{z-z \zbar}{z-\zbar},1\right)\right)+G(1,\zbar) \left(G\left(0,\frac{(z-1) \zbar}{z-\zbar},1\right)-G\left(0,-\frac{\zbar}{z-\zbar},1\right)+ \right.
	$$
	$$
	\left.G\left(0,\frac{\zbar-1}{\zbar-z},1\right)-G\left(1,\frac{(z-1) \zbar}{z-\zbar},1\right)+G\left(1,-\frac{\zbar}{z-\zbar},1\right)-G\left(1,\frac{\zbar-1}{\zbar-z},1\right)\right)+
	$$
	$$
	2 G\left(0,0,\frac{z-1}{z-\zbar},1\right)+2 G\left(0,0,\frac{\zbar-1}{\zbar-z},1\right)+2 G\left(0,1,\frac{z-1}{z-\zbar},1\right)+2 G\left(0,1,\frac{\zbar-1}{\zbar-z},1\right)+
	$$
	$$
	G\left(0,\frac{z-1}{z-\zbar},0,1\right)+G\left(0,\frac{z-1}{z-\zbar},1,1\right)+G\left(0,\frac{z-1}{z-\zbar},\frac{z-1}{z-\zbar},1\right)-G\left(0,\frac{z}{z-\zbar},\frac{z-1}{z-\zbar},1\right)-
	$$
	$$
	G\left(0,\frac{(z-1) \zbar}{z-\zbar},\frac{z-1}{z-\zbar},1\right)-G\left(0,\frac{z-z \zbar}{z-\zbar},\frac{\zbar-1}{\zbar-z},1\right)-G\left(0,-\frac{\zbar}{z-\zbar},\frac{\zbar-1}{\zbar-z},1\right)+
	$$
	$$
	G\left(0,\frac{\zbar-1}{\zbar-z},0,1\right)+G\left(0,\frac{\zbar-1}{\zbar-z},1,1\right)+G\left(0,\frac{\zbar-1}{\zbar-z},\frac{\zbar-1}{\zbar-z},1\right)-2 G\left(1,0,\frac{a-1}{a-b},1\right)-
	$$
	$$
	2 G\left(1,0,\frac{\zbar-1}{\zbar-z},1\right)-2 G\left(1,1,\frac{z-1}{z-\zbar},1\right)-2 G\left(1,1,\frac{\zbar-1}{\zbar-z},1\right)-G\left(1,\frac{z-1}{z-\zbar},0,1\right)-
	$$
	$$
	G\left(1,\frac{z-1}{z-\zbar},1,1\right)-G\left(1,\frac{z-1}{z-\zbar},\frac{z-1}{z-\zbar},1\right)+G\left(1,\frac{z}{z-\zbar},\frac{z-1}{z-\zbar},1\right)+G\left(1,\frac{(z-1) \zbar}{z-\zbar},\frac{z-1}{\zbar-\zbar},1\right)
	$$
	$$
	+G\left(1,\frac{z-z \zbar}{z-\zbar},\frac{z-1}{\zbar-z},1\right)+G\left(1,-\frac{\zbar}{z-\zbar},\frac{\zbar-1}{\zbar-z},1\right)-G\left(1,\frac{\zbar-1}{\zbar-z},0,1\right)-G\left(1,\frac{\zbar-1}{\zbar-z},1,1\right)
	$$
	\begin{equation}\label{firstcutfin}
		-G\left(1,\frac{\zbar-1}{\zbar-z},\frac{\zbar-1}{\zbar-z},1\right)
	\end{equation}
	Some of the terms in the above expression are divergent, and it turns out that the divergent parts from such terms exactly cancel each other, making the total contribution divergenceless. Note that this is a required criterion because a finite result shows that there will be no more divergent contributions in the form of more $\eps^{-1}$ terms. Similarly, the divergent parts of some terms from $f^{(-1)}(z,\zbar)$ cancels out, showing that there will no divergent contributions in the form of $\eps^{-2}$ terms.  The process where we show how these divergent parts cancel has been worked out explicitly in the Appendix for the total cut.
	%%%%%%%%%%%%%%%%%%%%%%%%%%%%%%%%%%%%%%%%%%%%%%%%%%%%%%%%%%%%%%%
	\section{The Second Cut}
	The Feynman integral for the second cut (see Fig.(\ref{scd})) is given by
	\beq
	\label{nonplanarsecondcut}
	\textrm{Cut}\hspace{0.05cm}\left[(k+l-p_2)^2,l^2,k^2\right]=- (2\pi)^3\,e^{2\gamma_E \eps}
	\int \frac{d^{D}l}{\pi^{D/2}}\frac{d^{D}k}{\pi^{D/2}}\, \frac{ \delta^+(l^2) \,\delta^+\left((k+l-p_2)^2\right)\,\delta^+\left(k^2\right)}{(k+l-p_1)^2 (k-p_1)^2(l-p_2)^2}\,,
	\eeq
	Again by inspecting the cut diagram we can write the integral as:
	$$
	\textrm{Cut}\hspace{0.05cm}\left[(k+l-p_2)^2,l^2,k^2\right]=
	$$
	\beq\label{secondcut}
	\frac{2\pi e^{\gamma_E \epsilon}}{\pi^{2-\epsilon}}\int d^{4-2\epsilon}l{\delta\left(l^2\right)\over (l-p_2)^2}
	\textrm{Cut}_{\left[(l-p_2)^2\right]}B^{3m}(0,p_3^2,(l-p_2)^2,p_1^2,s,t)\,.
	\eeq 
	where $	\textrm{Cut}_{\left[(l-p_2)^2\right]}B^{3m}(0,p_3^2,(l-p_2)^2,p_1^2,s,t)$ is the $p_3^2$ channel cut of a three mass box \\ $B^{3m}(0,p_2^2,p_3^2,p_4^2,s,t)$. \\
	\indent Now this can be evaluated in two ways - one is the direct way where we directly evaluate the integral with delta functions inside it in place of the propagators which are cut. It turns out that the calculations become very difficult using this method. The other way to evaluate is to use the Cutkosky's rule eq.(\ref{eq:oldcutting}) where we have to use the result for the three mass box from the literature and then find out the cut. The three mass box is given by the integral eq.(17) of \cite{Duplancic n nizic} 
	\begin{align}\label{threemassbox}
		B^{3m}(0,p_2^2,p_3^2,p_4^2,s,t) & =i\Gamma(1+\eps)\int_{0}^{1}dy~dz \frac{1}{z(s-p_2^2)+(1-z)(p_4^2-t)} \nonumber \\
		& \times \left\{\left[-y(1-y)(z~ s+(1-z)p_4^2)-z(1-z)y^2p_3^2\right]^{-1-\eps} \right. \nonumber \\ & \left.-\left[-y(1-y)(z~p_2^2+(1-z)t)-z(1-z)y^2p_3^2\right]^{-1-\eps}\right\}
	\end{align}
	The reason behind writing it in this way is that it becomes easier to evaluate the cut to all orders in $\eps$ using the above expression which is preferred over evaluating the cut of this box diagram when the result is only available upto $\eps^0$ order. Now the cut or the discontinuity across the $p_3^2$ channel can be evaluated using the following formula \cite{Abreu et al.(2015)}
	\begin{equation}\label{cutexponential}
		\text{Cut}_{a}[(a-b)^{-\eps}]=\text{Disc}_{a}[(a-b)^{-\eps}]=\frac{2\pi i \eps}{\Gamma(1-\eps)\Gamma(1+\eps)}(b-a)^{-\eps}\theta\left({a \over b}-1\right)
	\end{equation}
	Using this formula operated inside the integral we have 
	\begin{align}
		\text{Cut}&_{p_3^2}B^{3m}(0,p_2^2,p_3^2,p_4^2,s,t)  =-\frac{2\pi}{\Gamma(-\eps)}\int_{0}^{1}dz~ \frac{1}{z(s-p_2^2)+(1-z)(p_4^2-t)} \nonumber \\
		& \times \left\{\int_{0}^{z~ s+(1-z)p_4^2  \over  z~ s+(1-z)p_4^2+z(z-1)p_3^2}dy\left[y(1-y)(z~ s+(1-z)p_4^2)+z(1-z)y^2p_3^2\right]^{-1-\eps} \right. \nonumber \\ & \left.-\int_{0}^{z~p_2^2+(1-z)t  \over  z~p_2^2+(1-z)t+z(z-1)p_3^2}dy\left[y(1-y)(z~p_2^2+(1-z)t)+z(1-z)y^2p_3^2\right]^{-1-\eps}\right\}
	\end{align}
	which after some adjustments become
	\begin{align}
		&\text{Cut}_{p_3^2}B^{3m}(0,p_2^2,p_3^2,p_4^2,s,t)  =-\frac{2\pi}{\Gamma(1-\eps)}\int_{0}^{1}dz~ \frac{1}{z(s-p_2^2)+(1-z)(p_4^2-t)} \nonumber \\
		& \times \left\{{(z~ s+(1-z)p_4^2)^{-2\eps-1} \over  (z~ s+(1-z)p_4^2+z(z-1)p_3^2)^{-\eps} }~{}_2F_1\left(-\eps,1+\eps;1-\eps;1-{2z(z-1)p_3^2 \over z~ s+(1-z)p_4^2+z(z-1)p_3^2 }\right) \right. \nonumber \\ & \left. -{(z~p_2^2+(1-z)t)^{-2\eps-1} \over  (z~p_2^2+(1-z)t+z(z-1)p_3^2)^{-\eps} }~{}_2F_1\left(-\eps,1+\eps;1-\eps;1-{2z(z-1)p_3^2 \over z~p_2^2+(1-z)t+z(z-1)p_3^2 }\right)\right\}
	\end{align}
	This result is complete in all orders in $\eps$ which is actually required to evaluate the total contribution to the $\eps^0$ term for the second cut. Now we can use Euler's theorem for Hypergeometric functions namely
	\begin{equation}\label{Eulerth}
		{}_2F_1(a,b;c;x)=(1-x)^{c-a-b}{}_2F_1(c-a,c-b;c;x)
	\end{equation}
	to rewrite it as
	\begin{align}\label{cut3mb}
		&\text{Cut}_{p_3^2}B^{3m}(0,p_2^2,p_3^2,p_4^2,s,t)  =-\frac{2\pi}{\Gamma(1-\eps)}\int_{0}^{1}dz~ \frac{1}{z(s-p_2^2)+(1-z)(p_4^2-t)} \nonumber \\
		& \times \left\{{(z~ s+(1-z)p_4^2)^{-2\eps-1}(2z(z-1)p_3^2)^{-\eps} \over  (z~ s+(1-z)p_4^2+z(z-1)p_3^2)^{-2\eps} }~{}_2F_1\left(1,-2\eps;1-\eps;1-{2z(z-1)p_3^2 \over z~ s+(1-z)p_4^2+z(z-1)p_3^2 }\right) \right. \nonumber \\ & \left. -{(z~p_2^2+(1-z)t)^{-2\eps-1}(2z(z-1)p_3^2)^{-\eps} \over  (z~p_2^2+(1-z)t+z(z-1)p_3^2)^{-2\eps} }~{}_2F_1\left(1,-2\eps;1-\eps;1-{2z(z-1)p_3^2 \over z~p_2^2+(1-z)t+z(z-1)p_3^2 }\right)\right\}
	\end{align}
	The significance behind this step is that the prefactor term $(2p_3^2)^{-\eps}$ in the numerator actually helps us to cancel out the divergences of the remaining integration. We will discuss about it in more detail later. 
	\begin{comment}
	Now we can use partial fraction expansion to rewrite the denominator inside the integral in a convenient way 
	\begin{align}
	\text{Cut}&_{p_3^2}B^{3m}(0,p_2^2,p_3^2,p_4^2,s,t)  =-\frac{2\pi}{\Gamma(1-\eps)(s~t-p_2^2p_4^2)}\int_{0}^{1}dz\left[ \frac{s+t-p_2^2-p_4^2}{z(s-p_2^2)+(1-z)(p_4^2-t)}\right. \nonumber \\
	& \times \left\{{(z~ s+(1-z)p_4^2)^{-2\eps} \over  (z~ s+(1-z)p_4^2+z(z-1)p_3^2)^{-\eps} }~{}_2F_1\left(-\eps,1+\eps;1-\eps;1-{2z(z-1)p_3^2 \over z~ s+(1-z)p_4^2 }\right) \right. \nonumber \\ & \left. -{(z~p_2^2+(1-z)t)^{-2\eps} \over  (z~p_2^2+(1-z)t+z(z-1)p_3^2)^{-\eps} }~{}_2F_1\left(-\eps,1+\eps;1-\eps;1-{2z(z-1)p_3^2 \over z~p_2^2+(1-z)t }\right)\right\} \nonumber \\
	& - \left\{{(s-p_4^2)(z~ s+(1-z)p_4^2)^{-2\eps-1} \over  (z~ s+(1-z)p_4^2+z(z-1)p_3^2)^{-\eps} }~{}_2F_1\left(-\eps,1+\eps;1-\eps;1-{2z(z-1)p_3^2 \over z~ s+(1-z)p_4^2 }\right) \right. \nonumber \\ & \left.\left. -{(p_2^2-t)(z~p_2^2+(1-z)t)^{-2\eps-1} \over  (z~p_2^2+(1-z)t+z(z-1)p_3^2)^{-\eps} }~{}_2F_1\left(-\eps,1+\eps;1-\eps;1-{2z(z-1)p_3^2 \over z~p_2^2+(1-z)t }\right)\right\}\right]
	\end{align}
	\end{comment}
	
	Again we parametrize the variables as in eq.(\ref{cos})-eq.(\ref{lambda}), but this time there is only one delta function in the integral after explicit evaluation of the three mass box cut. This enforces $\abs{\bold{l}}=l_{0}$, but $l_0$ is left as an integration variable which we again parametrize as:
	\beq
	l_0\to{\sqrt{p_2^2}\over 2}y
	\eeq
	After which the momentum invariants become:
	$$s=p_1^2[(1-z)(1-\zbar)+y(z(1-\zbar)-x(z-\zbar))],\hspace{0.2cm}t=p_1^2[1-y(z-x(z-\zbar))],$$
	$$
	p_3^2=p_1^2[(1-z)(1-\bar z)],\hspace{0.2cm}p_2^2=p_1^2[z\bar z],\hspace{0.2cm}(l-p_2^2)^2=p_1^2(z\zbar)(1-y)
	$$
	and
	\begin{equation}\label{secin}
		s~ t-p_1^2p_3^2=(p_1^2)^2y[z^2(1-\zbar)+x(1-z)(1-\zbar)(z-\zbar)-y(z(1-\zbar)-x(z-\zbar))(z-x(z-\zbar))]
	\end{equation}
	and finally the momentum integration becomes
	\beq
	\frac{2\pi e^{\gamma_E \epsilon}}{\pi^{2-\epsilon}}\int d^{4-2\epsilon}l{\delta\left(l^2\right)\over (l-p_2)^2}=\frac{2e^{\gamma_E \epsilon}(z\bar{z})^{-\eps}(p_1^2)^{-\eps} }{\Gamma(1-\epsilon)}\int_{0}^{1}dx~ x^{-\eps}(1-x)^{-\eps}\int_{0}^{\infty}dy~y^{1-2\eps}(1-y)^{-1}
	\,.
	\eeq 
	Note that the $y$-integration limits are restricted to the positive real axis because of the Heaviside theta function in eq.(\ref{cutdefdiag})
	Substituting all this eq.(\ref{secondcut}) becomes
	$$
	\textrm{Cut}\hspace{0.05cm}\left[(k+l-p_2)^2,l^2,k^2\right]=\frac{4\pi c_{\Gamma}e^{\gamma_E \epsilon} }{\Gamma(1-\epsilon)}(z\bar{z})^{-\eps}(p_1^2)^{-2-2\eps}
	$$
	$$
	\times	(2z\zbar)^{-\eps}\int_{0}^{1}dx~ \int_{0}^{1}dy~x^{-\eps}(1-x)^{-\eps}y^{1-2\eps}(1-y)^{-1-\eps}\int_{0}^{1}{dz_1 \over c_{\Gamma}\Gamma(1-\eps)}\left[ {1 \over z_1(z\bar{z})-z+x(z-\bar{z})}\right.
	$$
	$$
	\times\left\{ {(z_1(z_1-1))^{-\eps}(z_1((1-z)(1-\bar{z})+y(z(1-\bar{z})-x(z-\zbar))-1)+1)^{-2\eps-1} \over (z_1((1-z)(1-\bar{z})+y(z(1-\bar{z})-x(z-\zbar))-1)+1+z_1(1-z_1)(z\zbar)(1-y))^{-2\eps} }\right. \times
	$$
	$$
	{}_2F_1\left(1,-2\eps;1-\eps;{z_1((1-z)(1-\bar{z})+y(z(1-\bar{z})-x(z-\zbar))-1)+1-z_1(1-z_1)(z\zbar)(1-y) \over z_1((1-z)(1-\bar{z})+y(z(1-\bar{z})-x(z-\zbar))-1)+1+z_1(1-z_1)(z\zbar)(1-y) }\right)
	$$
	$$
	-
	{(z_1(z_1-1))^{-\eps}(z_1((1-z)(1-\bar{z})-1+y(z-x(z-\zbar)))+1-y(z-x(z-\zbar)))^{-2\eps-1} \over (z_1((1-z)(1-\bar{z})-1+y(z-x(z-\zbar)))+1-y(z-x(z-\zbar))+z_1(1-z_1)(z\zbar)(1-y))^{-\eps} } 
	$$
	$$
	\times
	{}_2F_1\left(1,-2\eps;1-\eps;\right.
	$$
	\begin{align}\label{secondcutint}
		\left.\left.{z_1((1-z)(1-\bar{z})-1+y(z-x(z-\zbar)))+1-y(z-x(z-\zbar))-z_1(1-z_1)(z\zbar)(1-y) \over z_1((1-z)(1-\bar{z})-1+y(z-x(z-\zbar)))+1-y(z-x(z-\zbar))+z_1(1-z_1)(z\zbar)(1-y) }\right)\right\}
	\end{align}
	Note that the upper limit of the $y$- integration is changed to 1. This is because after evaluation of the $\text{Cut}_{p_3^2}B^{3m}(0,p_2^2,p_3^2,p_4^2,s,t)$ there is a $\theta(p_3^2)$ multiplied to the final result in eq.(\ref{cut3mb}) in accordance with eq.(\ref{cutdefdiag}). So we have in our case $(l-p_2^2)^2>0$ and hence using eq.(\ref{secin}) we have $y<1$.  Now the Hypergeometric ${}_2F_1$ function can be expanded order by order in $\eps$ using HypExp and the remaining integration can be done using PolyLogTools. \\
	\indent We can see that the second cut is not as trivial as the first cut. After the Hypergeometric ${}_2F_1$ expansion we have three integrations that need to be done in constrast to a single integration for the first cut. There are many contributions coming out to final answer. 
	\begin{equation}\label{secondcutf}
		\textrm{Cut}\hspace{0.05cm}\left[(k+l-p_2)^2,l^2,k^2\right]=\frac{4\pi c_{\Gamma}e^{\gamma_E \epsilon} }{\Gamma(1-\epsilon)}{(z\bar z)^{-\eps} \over (z-\bar z)^2}(p_1^2)^{-2-2\eps}\sum_{k=-1}^{\infty}\eps^k[g^{(k)}(z,\zbar)] 
	\end{equation}
	with 
	\begin{equation}
		g^{(-1)}(z,\zbar)=G\left(0,\frac{z-1}{z-\zbar},1\right)+G\left(0,\frac{\zbar-1}{\zbar-z},1\right)-G\left(1,\frac{z-1}{z-\zbar},1\right)-G\left(1,\frac{\zbar-1}{\zbar-z},1\right)
	\end{equation}
	and
	\begin{equation}
		g^{(0)}(z,\zbar)=g_M^{(0)}(z,\zbar)+g_{\eps^1}^{(0)}(z,\zbar)+g_{\eps^{-1}}^{(0)}(z,\zbar)+g_P^{(0)}(z,\zbar)
	\end{equation}
	We will describe each contribution as follows:
	\begin{flushleft}
		\begin{itemize}
			\item \textbf{The principal contribution :} 
		\end{itemize}
	\end{flushleft}
	After expanding the Hypergeometric ${}_2F_1$ functions in eq.(\ref{secondcutint}) in orders of epsilon, the obtained $\eps^0$ term gives the main contribution. After integration by $z_1$, $y$ and $x$, it is given by 
	$$
	g_M^{(0)}(z,\zbar)=
	$$
	$$
	G\left(\frac{z-1}{z-\zbar},1\right) \text{HPL}(\{1\},1)^2+G\left(\frac{\zbar-1}{\zbar-z},1\right) \text{HPL}(\{1\},1)^2+G(1,\zbar) G\left(\frac{z}{z-\zbar},1\right) \text{HPL}(\{1\},1)
	$$
	$$-G(1,\zbar) G\left(\frac{z-z \zbar}{z-\zbar},1\right) \text{HPL}(\{1\},1)+
	G\left(\frac{z-1}{z-\zbar},0,1\right) \text{HPL}(\{1\},1)+
	G\left(\frac{z-1}{z-\zbar},1,1\right) \text{HPL}(\{1\},1)
	$$
	$$-G\left(\frac{z-1}{z-\zbar},\frac{z-1}{z-\zbar},1\right) \text{HPL}(\{1\},1)+
	2 G\left(\frac{z}{z-\zbar},\frac{z-1}{z-\zbar},1\right) \text{HPL}(\{1\},1)+
	$$
	$$2 G\left(\frac{z-z \zbar}{z-\zbar},\frac{\zbar-1}{\zbar-z},1\right) \text{HPL}(\{1\},1)+
	G\left(\frac{\zbar-1}{\zbar-z},0,1\right) \text{HPL}(\{1\},1)+
	G\left(\frac{\zbar-1}{\zbar-z},1,1\right) \text{HPL}(\{1\},1)
	$$
	$$-G\left(\frac{\zbar-1}{\zbar-z},\frac{\zbar-1}{\zbar-z},1\right) \text{HPL}(\{1\},1)-G(1,\zbar) G\left(0,\frac{z}{z-\zbar},1\right)+
	G(1,\zbar) G\left(0,\frac{z-z \zbar
	}{z-\zbar},1\right)+
	$$
	$$G(1,\zbar) G\left(1,\frac{z}{z-\zbar},1\right)-G(1,\zbar) G\left(1,\frac{z-z \zbar}{z-\zbar},1\right)+G\left(0,0,\frac{z-1}{z-\zbar},1\right)+G\left(0,0,\frac{\zbar-1}{\zbar-z},1\right)+
	$$
	$$G\left(0,1,\frac{z-1}{z-\zbar},1\right)+G\left(0,1,\frac{\zbar-1}{\zbar-z},1\right)+G\left(0,\frac{z-1}{z-\zbar},\frac{z-1}{z-\zbar},1\right)-2 G\left(0,\frac{z}{z-\zbar},\frac{z-1}{z-\zbar},1\right)-
	$$
	$$2 G\left(0,\frac{z-z \zbar}{z-\zbar},\frac{\zbar-1}{\zbar-z},1\right)+G\left(0,\frac{\zbar-1}{\zbar-z},\frac{\zbar-1}{\zbar-z},1\right)-G\left(1,0,\frac{z-1}{z-\zbar},1\right)-G\left(1,0,\frac{\zbar-1}{\zbar-z},1\right)-
	$$
	$$
	G\left(1,1,\frac{z-1}{z-\zbar},1\right)-G\left(1,1,\frac{\zbar-1}{\zbar-z},1\right)-G\left(1,\frac{z-1}{z-\zbar},\frac{z-1}{z-\zbar},1\right)+2 G\left(1,\frac{z}{z-\zbar},\frac{z-1}{z-\zbar},1\right)+
	$$
	$$
	2 G\left(1,\frac{z-z \zbar}{z-\zbar},\frac{\zbar-1}{\zbar-z},1\right)-G\left(1,\frac{\zbar-1}{\zbar-z},\frac{\zbar-1}{\zbar-z},1\right)+G(0,(z-1) (\zbar-1)) \left(G\left(0,\frac{z}{z-\zbar},1\right)- \right.
	$$
	$$
	\left.G\left(0,\frac{z-z \zbar}{z-\zbar},1\right)-G\left(1,\frac{z}{z-\zbar},1\right)+G\left(1,\frac{z-z \zbar}{z-\zbar},1\right)-G\left(\frac{z}{z-\zbar},1\right) \text{HPL}(\{1\},1)+
	\right.
	$$
	$$
	\left.G\left(\frac{z-z \zbar}{z-\zbar},1\right) \text{HPL}(\{1\},1)\right)+G(1,z) \left(G\left(0,\frac{z-1}{z-\zbar},1\right)-3 G\left(0,\frac{z}{z-\zbar},1\right)+3 G\left(0,\frac{z-z \zbar}{z-\zbar},1\right)
	\right.
	$$
	$$
	\left.-G\left(0,\frac{\zbar-1}{\zbar-z},1\right)-G\left(1,\frac{z-1}{z-\zbar},1\right)+3 G\left(1,\frac{z}{z-\zbar},1\right)-3 G\left(1,\frac{z-z \zbar}{z-\zbar},1\right)+G\left(1,\frac{\zbar-1}{\zbar-z},1\right)
	\right.
	$$
	$$
	\left.-G\left(\frac{z-1}{z-\zbar},1\right) \text{HPL}(\{1\},1)+3 G\left(\frac{z}{z-\zbar},1\right) \text{HPL}(\{1\},1)-3 G\left(\frac{z-z \zbar}{z-\zbar},1\right) \text{HPL}(\{1\},1)
	\right.
	$$
	\begin{equation}
		\left.+G\left(\frac{\zbar-1}{\zbar-z},1\right) \text{HPL}(\{1\},1)\right)
	\end{equation}
	The final result again has many of the terms divergent, which exactly cancels out, giving the final result divergenceless.
	Here we have used the following formula to get rid of the end point singularities occuring for the $y$-integration at $y=1$.
	\begin{equation}\label{divcan}
		\int_{0}^{1}dy\frac{g(y,\eps)}{(1-y)^{1+\eps}}=\frac{g(1,\eps)}{\eps}+\int_{0}^{1}\frac{g(y,\eps)-g(1,\eps)}{(1-y)^{1+\eps}}
	\end{equation}
	
	Because of the first term in RHS of the above equation, this step produces the $\eps^{-1}$ term $g^{(-1)}(z,\zbar)$ and a subsequent contribution to the $\eps^{0}$ term    $g_{\eps^{-1}}^{(0)}(z,\zbar)$. Again we use a similar formula for the $x$-integration to get rid of the end-point singularities occurring at $x=0$ and $x=1$. The similar $\eps^{-1}$ kind of terms coming out after this step add up to zero and hence giving no contribution at all. Also we are able to use eq.(\ref{divcan}) to cancel the singularities only because of the step in eq.(\ref{Eulerth}) and eq.(\ref{cut3mb}).
	\begin{itemize}
		\item \textbf{The contribution from the $\eps^{-1}$ term:}
	\end{itemize}
	The $\eps^{-1}$ term $g^{(-1)}(z,\zbar)$ coming after the $y$-integration when multiplied with $x^{-\eps}(1-x)^{-\eps}$ gives a $\eps^0$ contribution which after the $x$-integration is given by
	$$
	g_{\eps^{-1}}^{(0)}(z,\zbar)=
	$$
	$$
	-G\left(0,0,\frac{z-1}{z-\bar{z}},1\right)-G\left(0,0,\frac{\bar{z}-1}{\bar{z}-z},1\right)-G\left(0,1,\frac{z-1}{z-\bar{z}},1\right)-G\left(0,1,\frac{\bar{z}-1}{\bar{z}-z},1\right)-G\left(0,\frac{z-1}{z-\bar{z}},0,1\right)
	$$
	$$-G\left(0,\frac{z-1}{z-\bar{z}},1,1\right)-G\left(0,\frac{\bar{z}-1}{\bar{z}-z},0,1\right)-G\left(0,\frac{\bar{z}-1}{\bar{z}-z},1,1\right)+G\left(1,0,\frac{z-1}{z-\bar{z}},1\right)+G\left(1,0,\frac{\bar{z}-1}{\bar{z}-z},1\right)
	$$
	$$+G\left(1,1,\frac{z-1}{z-\bar{z}},1\right)+G\left(1,1,\frac{\bar{z}-1}{\bar{z}-z},1\right)+G\left(1,\frac{z-1}{z-\bar{z}},0,1\right)+G\left(1,\frac{z-1}{z-\bar{z}},1,1\right)+G\left(1,\frac{\bar{z}-1}{\bar{z}-z},0,1\right)
	$$
	\begin{equation}
		+G\left(1,\frac{\bar{z}-1}{\bar{z}-z},1,1\right)
	\end{equation}
	Here also, the final result is divergenceless after the cancellation of the divergence arising from individual terms. Note that this is a kind of a feedback contribution that is arising only because of the need to cancel the singularities using eq.(\ref{divcan}) and does not have an existence of its own if there had not been any end-point singularity due to the $y$-integration at $y$=1. The divergences that occur due to the end-point singularities at $x=0$ and $x=1$ while doing $x$-integration add up to zero giving no net contribution.
	\begin{itemize}
		\item \textbf{The contribution from the $\eps^{1}$ term:}
	\end{itemize}
	After expanding the Hypergeometric ${}_2F_1$ functions in eq.(\ref{secondcutint}) in orders of epsilon, the obtained $\eps^1$ term gives back a contribution to the $\eps^0$ term since we need to get rid of the end-point singularities arising from the $y$-integration at $y=1$. It is given by
	$$
	g_{\eps^1}^{(0)}(z,\zbar)=
	$$
	$$
	-\left(2 G(1,z) \left(G\left(0,\frac{z-1}{z-\bar{z}},1\right)-G\left(0,\frac{z}{z-\bar{z}},1\right)+G\left(0,\frac{z-z \bar{z}}{z-\bar{z}},1\right)-G\left(1,\frac{z-1}{z-\bar{z}},1\right)+
	\right.\right.
	$$
	$$
	\left.G\left(1,\frac{z}{z-\bar{z}},1\right)-G\left(1,\frac{z-z \bar{z}}{z-\bar{z}},1\right)\right)+G\left(1,\bar{z}\right) \left(G\left(0,\frac{z-1}{z-\bar{z}},1\right)+G\left(0,\frac{\bar{z}-1}{\bar{z}-z},1\right)-
	\right.
	$$
	$$
	\left.G\left(1,\frac{z-1}{z-\bar{z}},1\right)-G\left(1,\frac{\bar{z}-1}{\bar{z}-z},1\right)\right)+2 G\left(0,0,\frac{z-1}{z-\bar{z}},1\right)+2 G\left(0,0,\frac{\bar{z}-1}{\bar{z}-z},1\right)+
	$$
	$$2 G\left(0,1,\frac{z-1}{z-\bar{z}},1\right)+2 G\left(0,1,\frac{\bar{z}-1}{\bar{z}-z},1\right)+G\left(0,\frac{z-1}{z-\bar{z}},\frac{z-1}{z-\bar{z}},1\right)-2 G\left(0,\frac{z}{z-\bar{z}},\frac{z-1}{z-\bar{z}},1\right)-
	$$
	$$2 G\left(0,\frac{z-z \bar{z}}{z-\bar{z}},\frac{\bar{z}-1}{\bar{z}-z},1\right)+G\left(0,\frac{\bar{z}-1}{\bar{z}-z},\frac{\bar{z}-1}{\bar{z}-z},1\right)-2 G\left(1,0,\frac{z-1}{z-\bar{z}},1\right)-2 G\left(1,0,\frac{\bar{z}-1}{\bar{z}-z},1\right)-
	$$
	$$
	2 G\left(1,1,\frac{z-1}{z-\bar{z}},1\right)-2 G\left(1,1,\frac{\bar{z}-1}{\bar{z}-z},1\right)-G\left(1,\frac{z-1}{z-\bar{z}},\frac{z-1}{z-\bar{z}},1\right)+2 G\left(1,\frac{z}{z-\bar{z}},\frac{z-1}{z-\bar{z}},1\right)+
	$$
	\begin{equation}
		\left.2 G\left(1,\frac{z-z \bar{z}}{z-\bar{z}},\frac{\bar{z}-1}{\bar{z}-z},1\right)-G\left(1,\frac{\bar{z}-1}{\bar{z}-z},\frac{\bar{z}-1}{\bar{z}-z},1\right)\right)
	\end{equation}
	Again the divergent parts cancel here. This term is again a feedback term that is produced only due to the end-point singularity at $y$=1. Here also the divergences due to the end-point singularities of $x$-integration add up to zero. Note that as discussed earlier, the contribution due to this term could not have been calculated if we had taken the result from literature for eq.(\ref{threemassbox}) only up to $\eps^0$ term.
	\begin{itemize}
		\item \textbf{The contribution from the pre-factor term:}
	\end{itemize}
	There is an additional prefactor term $(2z\zbar)^{-\eps}$ in eq.(\ref{secondcutint}) which contributes to the overall $\eps^0$ term through its product with the $\eps^{-1}$ term $g^{(-1)}(z,\zbar)$. It is given by
	$$
	g_P^{(0)}(z,\zbar)=
	$$
	\begin{equation}\label{pre-factor}
		\left(G\left(0,\bar{z}\right)+G(0,z)\right) \left(G\left(0,\frac{z-1}{z-\bar{z}},1\right)+G\left(0,\frac{\bar{z}-1}{\bar{z}-z},1\right)-G\left(1,\frac{z-1}{z-\bar{z}},1\right)-G\left(1,\frac{\bar{z}-1}{\bar{z}-z},1\right)\right)
	\end{equation}
	Note that for each contribution, the divergences inside them always cancel out. This has been explicitly checked for each contribution. This term is again produced only because of the steps in eq.(\ref{Eulerth}) and eq.(\ref{cut3mb}) to cancel the singularities. \\
	\indent The last three contributions are not at all trivial to guess at first in the sense that the prefactor, the lower order ($\eps^{-1}$) terms, and higher-order ($\eps^{1}$) terms contribute back to the required result.
	%%%%%%%%%%%%%%%%%%%%%%%%%%%%%%%%%%%%%%%%%%%%%%%%%%%%%%%%%%%%%%%%%%%%%
	\section{Summing of the cuts}
	Now we have all the contributions from both the cuts and hence we can find the net cut across the $p_2^2$ channel, which is a sum of the two cuts. Since the prefactors of both the cuts match, we can just focus on the multiple polylogarithms part. As expected, since the original Feynman integral is not divergent in 4 dimensions, the divergences ($\eps^{-1}$ terms) cancel:
	\begin{equation}\label{totalcutdiv}
		f^{(-1)}(z,\zbar)+g^{(-1)}(z,\zbar)=0
	\end{equation}
	Again as expected, the $\eps^0$ (finite) terms do not cancel, and they give the following result.
	$$
	f^{(0)}(z,\zbar)+	g^{(0)}(z,\zbar)=
	$$
	$$
	-G\left(1,\bar{z}\right) G\left(0,\frac{z-1}{z-\bar{z}},1\right)+G\left(0,(z-1) \left(\bar{z}-1\right)\right) G\left(0,\frac{z}{z-\bar{z}},1\right)-2 G(1,z) G\left(0,\frac{z}{z-\bar{z}},1\right)-
	$$
	$$G\left(1,\bar{z}\right) G\left(0,\frac{z}{z-\bar{z}},1\right)+G\left(1,\bar{z}\right) G\left(0,\frac{(z-1) \bar{z}}{z-\bar{z}},1\right)-G\left(0,(z-1) \left(\bar{z}-1\right)\right) G\left(0,\frac{z-z \bar{z}}{z-\bar{z}},1\right)+
	$$
	$$
	2 G(1,z) G\left(0,\frac{z-z \bar{z}}{z-\bar{z}},1\right)+G\left(1,\bar{z}\right) G\left(0,\frac{z-z \bar{z}}{z-\bar{z}},1\right)-G\left(1,\bar{z}\right) G\left(0,-\frac{\bar{z}}{z-\bar{z}},1\right)-
	$$
	$$
	G(1,z) G\left(0,\frac{\bar{z}-1}{\bar{z}-z},1\right)-G\left(1,\bar{z}\right) G\left(\frac{z-1}{z-\bar{z}},1,1\right)+G\left(0,(z-1) \left(\bar{z}-1\right)\right) G\left(\frac{z}{z-\bar{z}},1,1\right)-
	$$
	$$
	2 G(1,z) G\left(\frac{z}{z-\bar{z}},1,1\right)-G\left(1,\bar{z}\right) G\left(\frac{z}{z-\bar{z}},1,1\right)+G\left(1,\bar{z}\right) G\left(\frac{(z-1) \bar{z}}{z-\bar{z}},1,1\right)-
	$$
	$$
	G\left(0,(z-1) \left(\bar{z}-1\right)\right) G\left(\frac{z-z \bar{z}}{z-\bar{z}},1,1\right)+2 G(1,z) G\left(\frac{z-z \bar{z}}{z-\bar{z}},1,1\right)+G\left(1,\bar{z}\right) G\left(\frac{z-z \bar{z}}{z-\bar{z}},1,1\right)-
	$$
	$$
	G\left(1,\bar{z}\right) G\left(-\frac{\bar{z}}{z-\bar{z}},1,1\right)-G(1,z) G\left(\frac{\bar{z}-1}{\bar{z}-z},1,1\right)+G(0,z) \left(G\left(0,\frac{z-1}{z-\bar{z}},1\right)+G\left(0,\frac{\bar{z}-1}{\bar{z}-z},1\right)
	\right.
	$$
	$$
	\left.+G\left(\frac{z-1}{z-\bar{z}},1,1\right)+G\left(\frac{\bar{z}-1}{\bar{z}-z},1,1\right)\right)+G\left(0,\bar{z}\right) \left(G\left(0,\frac{z-1}{z-\bar{z}},1\right)+G\left(0,\frac{\bar{z}-1}{\bar{z}-z},1\right)+\right.
	$$
	$$
	\left.G\left(\frac{z-1}{z-\bar{z}},1,1\right)+G\left(\frac{\bar{z}-1}{\bar{z}-z},1,1\right)\right)+G\left(0,\frac{z-1}{z-\bar{z}},\frac{z-1}{z-\bar{z}},1\right)-G\left(0,\frac{z}{z-\bar{z}},\frac{z-1}{z-\bar{z}},1\right)-
	$$
	$$
	G\left(0,\frac{(z-1) \bar{z}}{z-\bar{z}},\frac{z-1}{z-\bar{z}},1\right)-G\left(0,\frac{z-z \bar{z}}{z-\bar{z}},\frac{\bar{z}-1}{\bar{z}-z},1\right)-G\left(0,-\frac{\bar{z}}{z-\bar{z}},\frac{\bar{z}-1}{\bar{z}-z},1\right)+
	$$
	$$
	G\left(0,\frac{\bar{z}-1}{\bar{z}-z},\frac{\bar{z}-1}{\bar{z}-z},1\right)+G\left(\frac{z-1}{z-\bar{z}},1,\frac{z-1}{z-\bar{z}},1\right)+G\left(\frac{z-1}{z-\bar{z}},\frac{z-1}{z-\bar{z}},1,1\right)-G\left(\frac{z}{z-\bar{z}},1,\frac{z-1}{z-\bar{z}},1\right)
	$$
	$$
	-G\left(\frac{z}{z-\bar{z}},\frac{z-1}{z-\bar{z}},1,1\right)-G\left(\frac{(z-1) \bar{z}}{z-\bar{z}},1,\frac{z-1}{z-\bar{z}},1\right)-G\left(\frac{(z-1) \bar{z}}{z-\bar{z}},\frac{z-1}{z-\bar{z}},1,1\right)-
	$$
	$$
	G\left(\frac{z-z \bar{z}}{z-\bar{z}},1,\frac{\bar{z}-1}{\bar{z}-z},1\right)-G\left(\frac{z-z \bar{z}}{z-\bar{z}},\frac{\bar{z}-1}{\bar{z}-z},1,1\right)-G\left(-\frac{\bar{z}}{z-\bar{z}},1,\frac{\bar{z}-1}{\bar{z}-z},1\right)-
	$$
	\begin{equation}\label{finalcut}
		G\left(-\frac{\bar{z}}{z-\bar{z}},\frac{\bar{z}-1}{\bar{z}-z},1,1\right)+G\left(\frac{\bar{z}-1}{\bar{z}-z},1,\frac{\bar{z}-1}{\bar{z}-z},1\right)+G\left(\frac{\bar{z}-1}{\bar{z}-z},\frac{\bar{z}-1}{\bar{z}-z},1,1\right)
	\end{equation}
	Here we have used the command \textbf{ShuffleRegulate} of PolyLogTools to cancel the divergent terms within themselves and rewrite the whole result with each term being non-divergent (see Appendix). 
	
	%%%%%%%%%%%%%%%%%%%%%%%%%%%%%%%%%%%%%%%%%%%%%%%%%%%%%%%%%%%%%%%%%%%
	\section{Reconstruction of the Symbol of the original Feynman Integral}
	As discussed earlier, the original Feynman integral can be reconstructed using the symbol $\cS(F)$ which is equal to the maximal iteration of the coproduct corresponding to the partition (1,...,1) $\Delta_{1,...,1}(F)$. Thus, for getting this, we will first find out the maximal iteration of the coproduct for our obtained result eq.(\ref{finalcut}). This can be done using the commands \textbf{SymbolExpand} and \textbf{SymbolMap} of PolyLogTools, and we get the following result:
	$$
	-2 \left(-(1-z)\otimes (1-z)\otimes \bar{z}-(1-z)\otimes \bar{z}\otimes (1-z)+(1-z)\otimes \bar{z}\otimes \left(1-\bar{z}\right)+z\otimes (1-z)\otimes \left(1-\bar{z}\right)\right.
	$$
	$$
	+z\otimes \left(1-\bar{z}\right)\otimes (1-z)-z\otimes \left(1-\bar{z}\right)\otimes \left(1-\bar{z}\right)+\left(1-\bar{z}\right)\otimes z\otimes (1-z)-\left(1-\bar{z}\right)\otimes z\otimes \left(1-\bar{z}\right)
	$$
	$$
	-\left(1-\bar{z}\right)\otimes \left(1-\bar{z}\right)\otimes z+\left(1-\bar{z}\right)\otimes \left(1-\bar{z}\right)\otimes \bar{z}-\bar{z}\otimes (1-z)\otimes (1-z)+\bar{z}\otimes (1-z)\otimes \left(1-\bar{z}\right)
	$$
	\begin{equation}
		\left.+\bar{z}\otimes \left(1-\bar{z}\right)\otimes (1-z)-\bar{z}\otimes \left(1-\bar{z}\right)\otimes \left(1-\bar{z}\right)+(1-z)\otimes (1-z)\otimes z-z\otimes (1-z)\otimes (1-z)\right)
	\end{equation}
	
	Now we have to apply the first entry condition and the integrability condition in order to get the  $\Delta_{1,...,1}(F)$. Since it is the $p_2^2$ channel cut we need to have $z\zbar$ as the first entry to the above expression:
	$$
	(z\zbar)\otimes[-2 \left(-(1-z)\otimes (1-z)\otimes \bar{z}-(1-z)\otimes \bar{z}\otimes (1-z)+(1-z)\otimes \bar{z}\otimes \left(1-\bar{z}\right)+\right.
	$$
	$$
	z\otimes (1-z)\otimes \left(1-\bar{z}\right)+z\otimes \left(1-\bar{z}\right)\otimes (1-z)-z\otimes \left(1-\bar{z}\right)\otimes \left(1-\bar{z}\right)+\left(1-\bar{z}\right)\otimes z\otimes (1-z)-
	$$
	$$
	\left(1-\bar{z}\right)\otimes z\otimes \left(1-\bar{z}\right)-\left(1-\bar{z}\right)\otimes \left(1-\bar{z}\right)\otimes z+\left(1-\bar{z}\right)\otimes \left(1-\bar{z}\right)\otimes \bar{z}-\bar{z}\otimes (1-z)\otimes (1-z)
	$$
	$$
	+\bar{z}\otimes (1-z)\otimes \left(1-\bar{z}\right)+\bar{z}\otimes \left(1-\bar{z}\right)\otimes (1-z)-\bar{z}\otimes \left(1-\bar{z}\right)\otimes \left(1-\bar{z}\right)+(1-z)\otimes (1-z)\otimes z
	$$
	\begin{equation}
		\left.-z\otimes (1-z)\otimes (1-z)\right)]
	\end{equation}
	
	Now in order to satisfy the integrability condition and the first entry condition, we add the following terms to this expression.
	$$
	2 \left((1-z)\otimes (1-z)\otimes z\otimes z-(1-z)\otimes (1-z)\otimes z\otimes \bar{z}-(1-z)\otimes (1-z)\otimes \bar{z}\otimes z+\right.
	$$
	$$
	(1-z)\otimes (1-z)\otimes \bar{z}\otimes \bar{z}-
	(1-z)\otimes z\otimes z\otimes (1-z)+(1-z)\otimes z\otimes z\otimes \left(1-\bar{z}\right)+
	$$
	$$
	(1-z)\otimes z\otimes \left(1-\bar{z}\right)\otimes z-(1-z)\otimes z\otimes \left(1-\bar{z}\right)\otimes \bar{z}+(1-z)\otimes \left(1-\bar{z}\right)\otimes z\otimes z-
	$$
	$$
	(1-z)\otimes \left(1-\bar{z}\right)\otimes z\otimes \bar{z}-(1-z)\otimes \left(1-\bar{z}\right)\otimes \bar{z}\otimes z+(1-z)\otimes \left(1-\bar{z}\right)\otimes \bar{z}\otimes \bar{z}-
	$$
	$$
	(1-z)\otimes \bar{z}\otimes (1-z)\otimes z+(1-z)\otimes \bar{z}\otimes (1-z)\otimes \bar{z}+(1-z)\otimes \bar{z}\otimes \bar{z}\otimes (1-z)-
	$$
	$$
	(1-z)\otimes \bar{z}\otimes \bar{z}\otimes \left(1-\bar{z}\right)-z\otimes (1-z)\otimes (1-z)\otimes z+z\otimes (1-z)\otimes (1-z)\otimes \bar{z}+
	$$
	$$
	\left(1-\bar{z}\right)\otimes (1-z)\otimes z\otimes \bar{z}-\left(1-\bar{z}\right)\otimes (1-z)\otimes \bar{z}\otimes z+\left(1-\bar{z}\right)\otimes (1-z)\otimes \bar{z}\otimes \bar{z}-
	$$
	$$
	\left(1-\bar{z}\right)\otimes z\otimes z\otimes (1-z)+\left(1-\bar{z}\right)\otimes z\otimes z\otimes \left(1-\bar{z}\right)+\left(1-\bar{z}\right)\otimes z\otimes \left(1-\bar{z}\right)\otimes z-
	$$
	$$
	\left(1-\bar{z}\right)\otimes z\otimes \left(1-\bar{z}\right)\otimes \bar{z}+\left(1-\bar{z}\right)\otimes \left(1-\bar{z}\right)\otimes z\otimes z-\left(1-\bar{z}\right)\otimes \left(1-\bar{z}\right)\otimes z\otimes \bar{z}-
	$$
	$$
	\left(1-\bar{z}\right)\otimes \left(1-\bar{z}\right)\otimes \bar{z}\otimes z+\left(1-\bar{z}\right)\otimes \left(1-\bar{z}\right)\otimes \bar{z}\otimes \bar{z}-\left(1-\bar{z}\right)\otimes \bar{z}\otimes (1-z)\otimes z+
	$$
	\begin{equation}\label{symbol}
		\left(1-\bar{z}\right)\otimes \bar{z}\otimes (1-z)\otimes \bar{z}+\left(1-\bar{z}\right)\otimes \bar{z}\otimes \bar{z}\otimes (1-z)-\left(1-\bar{z}\right)\otimes \bar{z}\otimes \bar{z}\otimes \left(1-\bar{z}\right)
	\end{equation}
	Finally adding the above two expressions we get
	$$
	2 \left((1-z)\otimes (1-z)\otimes z\otimes z-(1-z)\otimes (1-z)\otimes z\otimes \bar{z}-(1-z)\otimes (1-z)\otimes \bar{z}\otimes z+\right.
	$$
	$$
	(1-z)\otimes (1-z)\otimes \bar{z}\otimes \bar{z}-
	(1-z)\otimes z\otimes z\otimes (1-z)+(1-z)\otimes z\otimes z\otimes \left(1-\bar{z}\right)+
	$$
	$$
	(1-z)\otimes z\otimes \left(1-\bar{z}\right)\otimes z-(1-z)\otimes z\otimes \left(1-\bar{z}\right)\otimes \bar{z}+(1-z)\otimes \left(1-\bar{z}\right)\otimes z\otimes z-
	$$
	$$
	(1-z)\otimes \left(1-\bar{z}\right)\otimes z\otimes \bar{z}-(1-z)\otimes \left(1-\bar{z}\right)\otimes \bar{z}\otimes z+(1-z)\otimes \left(1-\bar{z}\right)\otimes \bar{z}\otimes \bar{z}-
	$$
	$$
	(1-z)\otimes \bar{z}\otimes (1-z)\otimes z+(1-z)\otimes \bar{z}\otimes (1-z)\otimes \bar{z}+(1-z)\otimes \bar{z}\otimes \bar{z}\otimes (1-z)-
	$$
	$$
	(1-z)\otimes \bar{z}\otimes \bar{z}\otimes \left(1-\bar{z}\right)-z\otimes (1-z)\otimes (1-z)\otimes z+z\otimes (1-z)\otimes (1-z)\otimes \bar{z}+
	$$
	$$z\otimes (1-z)\otimes \bar{z}\otimes (1-z)-
	z\otimes (1-z)\otimes \bar{z}\otimes \left(1-\bar{z}\right)+z\otimes z\otimes (1-z)\otimes (1-z)-
	$$
	$$
	z\otimes z\otimes (1-z)\otimes \left(1-\bar{z}\right)-z\otimes z\otimes \left(1-\bar{z}\right)\otimes (1-z)+z\otimes z\otimes \left(1-\bar{z}\right)\otimes \left(1-\bar{z}\right)-
	$$
	$$
	z\otimes \left(1-\bar{z}\right)\otimes z\otimes (1-z)+z\otimes \left(1-\bar{z}\right)\otimes z\otimes \left(1-\bar{z}\right)+z\otimes \left(1-\bar{z}\right)\otimes \left(1-\bar{z}\right)\otimes z-
	$$
	$$
	z\otimes \left(1-\bar{z}\right)\otimes \left(1-\bar{z}\right)\otimes \bar{z}+z\otimes \bar{z}\otimes (1-z)\otimes (1-z)-z\otimes \bar{z}\otimes (1-z)\otimes \left(1-\bar{z}\right)-
	$$
	$$
	z\otimes \bar{z}\otimes \left(1-\bar{z}\right)\otimes (1-z)+z\otimes \bar{z}\otimes \left(1-\bar{z}\right)\otimes \left(1-\bar{z}\right)+\left(1-\bar{z}\right)\otimes (1-z)\otimes z\otimes z-
	$$
	$$
	\left(1-\bar{z}\right)\otimes (1-z)\otimes z\otimes \bar{z}-\left(1-\bar{z}\right)\otimes (1-z)\otimes \bar{z}\otimes z+\left(1-\bar{z}\right)\otimes (1-z)\otimes \bar{z}\otimes \bar{z}-
	$$
	$$
	\left(1-\bar{z}\right)\otimes z\otimes z\otimes (1-z)+\left(1-\bar{z}\right)\otimes z\otimes z\otimes \left(1-\bar{z}\right)+\left(1-\bar{z}\right)\otimes z\otimes \left(1-\bar{z}\right)\otimes z-
	$$
	$$
	\left(1-\bar{z}\right)\otimes z\otimes \left(1-\bar{z}\right)\otimes \bar{z}+\left(1-\bar{z}\right)\otimes \left(1-\bar{z}\right)\otimes z\otimes z-\left(1-\bar{z}\right)\otimes \left(1-\bar{z}\right)\otimes z\otimes \bar{z}-
	$$
	$$
	\left(1-\bar{z}\right)\otimes \left(1-\bar{z}\right)\otimes \bar{z}\otimes z+\left(1-\bar{z}\right)\otimes \left(1-\bar{z}\right)\otimes \bar{z}\otimes \bar{z}-\left(1-\bar{z}\right)\otimes \bar{z}\otimes (1-z)\otimes z+
	$$
	$$
	\left(1-\bar{z}\right)\otimes \bar{z}\otimes (1-z)\otimes \bar{z}+\left(1-\bar{z}\right)\otimes \bar{z}\otimes \bar{z}\otimes (1-z)-\left(1-\bar{z}\right)\otimes \bar{z}\otimes \bar{z}\otimes \left(1-\bar{z}\right)-
	$$
	$$
	\bar{z}\otimes (1-z)\otimes (1-z)\otimes z+\bar{z}\otimes (1-z)\otimes (1-z)\otimes \bar{z}+\bar{z}\otimes (1-z)\otimes \bar{z}\otimes (1-z)-
	$$
	$$
	\bar{z}\otimes (1-z)\otimes \bar{z}\otimes \left(1-\bar{z}\right)+
	\bar{z}\otimes z\otimes (1-z)\otimes (1-z)-\bar{z}\otimes z\otimes (1-z)\otimes \left(1-\bar{z}\right)-
	$$
	$$
	\bar{z}\otimes z\otimes \left(1-\bar{z}\right)\otimes (1-z)+\bar{z}\otimes z\otimes \left(1-\bar{z}\right)\otimes \left(1-\bar{z}\right)-\bar{z}\otimes \left(1-\bar{z}\right)\otimes z\otimes (1-z)+
	$$
	$$
	\bar{z}\otimes \left(1-\bar{z}\right)\otimes z\otimes \left(1-\bar{z}\right)+\bar{z}\otimes \left(1-\bar{z}\right)\otimes \left(1-\bar{z}\right)\otimes z-\bar{z}\otimes \left(1-\bar{z}\right)\otimes \left(1-\bar{z}\right)\otimes \bar{z}+
	$$
	$$
	\bar{z}\otimes \bar{z}\otimes (1-z)\otimes (1-z)-\bar{z}\otimes \bar{z}\otimes (1-z)\otimes \left(1-\bar{z}\right)-\bar{z}\otimes \bar{z}\otimes \left(1-\bar{z}\right)\otimes (1-z)+
	$$
	\begin{equation}\label{symbol1}
		\left.
		\bar{z}\otimes \bar{z}\otimes \left(1-\bar{z}\right)\otimes \left(1-\bar{z}\right)\right)
	\end{equation}
	At this point, it satisfies both the integrability and the first entry condition. Also,
	this coincides with the coproduct of the original Feynman integral, eq.(\ref{orgfeyn}) when it is written in terms of the variables $z$ and $\zbar$.
	$$C(p_1^2,p_2^2,p_3^3)=$$
	\begin{equation}\label{originalanalyticresult}
		\frac{\left(2 \text{Li}_2\left(\frac{1}{\bar{z}}\right)-\log (-z) \log \left(-\bar{z}\right)+\left(\log \left(z\bar{z}\right)\right) \left(\log \left(\bar{z}\right)+\log(-{1-z \over 1-\zbar}) \right)+2 \text{Li}_2(z)+\frac{\pi ^2}{3}\right)^2}{(z-\zbar)^2}
	\end{equation}
	We can reconstruct this function using the obtained coproduct by predicting the family of functions that gives this particular coproduct and then extract out that unique function out of the family of the functions by using the rule
	\begin{equation}\label{cutkorule}
		\Disc_{p_2^2}C(p_1^2,p_2^2,p_3^3)\cong\Cut_{p_2^2}C(p_1^2,p_2^2,p_3^3)
	\end{equation}
	This Method is well described in Section 7 of \cite{Abreu et al.(2014)}. Thus in this section we have shown the consistency of the computation of the symbol obtained from the known two loop Feynman integral computation. Eq.(\ref{symbol1}) is the one of the main results of this paper which verifies the Duhr conjecture \cite{Duhr:2012fh}.
	
	\section{Reconstruction of the full function}
	The symbol alphabet for our non-planar case is same as the massless triangle and massless two-loop ladder. It is given by $\cA=\{z,\zbar,1-z,1-\zbar\}$. We will use the fact that the most general class of functions giving rise to this symbol alphabet and satisfying the first entry condition are the single-valued harmonic polylogarithms \cite{BrownSVHPLs}. 
	
	In order to achieve this goal, we look at each term of the obtained symbol eq.(\ref{symbol1}). We need to examine the entire set of HPl's of weight less than 4 which can give rise to the symbol using the commands \textbf{SymbolExpand} and \textbf{SymbolMap} of PolyLogTools. For example the symbol generated by the HPL $H(0,0,1,1,z)$ is the first term $(1-z)\otimes (1-z)\otimes z\otimes z$. We can clearly see the pattern of the indices containing 0's and 1's which gives rise to the required symbol. Therefore it may be seen that the index $1$ gives rise to terms of $(1-z)$ kind and index $0$ gives rise to terms of $(z)$ kind and the index $-1$ gives rise to $(1+z)$ kind which does not appear in the whole symbol and hence the HPL's containing index $-1$ do not contribute to the final result. This pattern follows from the structure of the HPL's and their associated symbols.
	
	For the terms which contain both the varables $z$ and $\bar{z}$ we can only have product HPL's of lower weight containing the two variables contributing them. The total weight of the product terms is 4 as required. Here also the arguments are such that they give rise to particular kind of symbol. For example,
	the product HPL $H(0,1,1,z) H\left(0,\bar{z}\right)$ gives rise to the following symbol:
	$$(1-z)\otimes (1-z)\otimes z\otimes \bar{z}+(1-z)\otimes (1-z)\otimes \bar{z}\otimes z+(1-z)\otimes \bar{z}\otimes (1-z)\otimes z+\bar{z}\otimes (1-z)\otimes (1-z)\otimes z$$
	Again here we have only $(\bar{z})$ inside the symbol and no $(1-\bar{z})$ and hence according to the pattern observed above, we have the index 0 inside the HPL with variable $\bar{z}$. Also all of these four terms are present inside our symbol eq.(\ref{symbol1}) and hence the product HPL  $H(0,1,1,z) H\left(0,\bar{z}\right)$ contributes to our symbol. Its existence as a contributor to our symbol is predicted by looking at the second term inside the symbol $(1-z)\otimes (1-z)\otimes z\otimes \bar{z}$ and comparing with the observed pattern. Likewise we can find out all the HPL's which give rise to the whole symbol by looking at the form of the remaining terms in the symbol which are not yet covered by any HPL's. Using this procedure we get the total sum of all such HPL's which is given by 
	$$4 \left(-H(0,1,z) H\left(0,1,\bar{z}\right)-H(1,0,z) H\left(1,0,\bar{z}\right)+H(1,1,z) H\left(0,0,\bar{z}\right)+H(0,0,z) H\left(1,1,\bar{z}\right)+\right.$$ $$\left.H(0,0,1,z) H\left(1,\bar{z}\right)+H(1,z) H\left(0,0,1,\bar{z}\right)-H(0,1,1,z) H\left(0,\bar{z}\right)-H(0,z) H\left(0,1,1,\bar{z}\right)-\right.$$ $$\left.H(1,0,0,z) H\left(1,\bar{z}\right)-H(1,z) H\left(1,0,0,\bar{z}\right)+H(1,1,0,z) H\left(0,\bar{z}\right)+H(0,z) H\left(1,1,0,\bar{z}\right)+\right.$$ $$\left.H\left(0,0,1,1,\bar{z}\right)-H\left(0,1,1,0,\bar{z}\right)-H\left(1,0,0,1,\bar{z}\right)+H\left(1,1,0,0,\bar{z}\right)+H(0,0,1,1,z)-\right.$$ 
	\begin{equation}\label{fullfunc}
		\left.H(0,1,1,0,z)-H(1,0,0,1,z)+H(1,1,0,0,z)\right) + \sum_ic_iH_i(z,\bar{z})
	\end{equation}
	Here $H_i(z,\bar{z})$ are the lower weight HPL's that do not contribute to the symbol eq.(\ref{symbol1}) but contribute to the final full function and the $c_i$ are coefficients which also carry a weight such that each term in the sum has total weight 4. The total sum $\sum_ic_iH_i(z,\bar{z})$ consists of a basis of weight 3,2,1 and 0 HPL's such that any HPL of respective weight can be represented by a linear combination of the basis HPL's with suitable constants $c_i$.

	Note that for the functions of pure type i.e. which are not product of two HPL's there may be product of two lower weight HPL's giving rise to the same symbol but since using the shuffle product of HPL's we can rewrite the product HPL's in terms of pure functions, therefore the pure functions form a basis and hence it is enough to represent it in the present form. This function when written in terms of classical polylogarithms by first using the command HToHPL of PolyLogTools and then using HPLConvertToKnownFunctions command of HPL Package is equal to the following expression:
	$$\left(-2 \text{Li}_2\left(\bar{z}\right)-\log ^2\left(\bar{z}\right)-\log (z) \log \left(\bar{z}\right)+\left(\log \left(\bar{z}\right)+\log (z)\right) \left(\log({1-z \over 1-\zbar})+\log \left(\bar{z}\right)\right)+2 \text{Li}_2(z)\right)^2 $$
	\begin{equation}\label{signflip}
		+ \sum_ic_iH_i(z,\bar{z})
	\end{equation}
	This result has to be further treated in order to get the required correspondence with eq.(\ref{originalanalyticresult}), which reads after employing the dilogarithm identities as
	$$\left(-2 \text{Li}_2\left(\bar{z}\right)-\log ^2\left(-\bar{z}\right)-\log (-z) \log \left(-\bar{z}\right)+\left(\log \left(z\bar{z}\right)\right) \left(\log(-{1-z \over 1-\zbar})+\log \left(\bar{z}\right)\right)+2 \text{Li}_2(z)\right)^2 $$.
	
	The expression above differs from the known result in the signs of the arguments of the simple logarithms. This is not worrisome, because the symbol method alone does not fix those signs in eq.(\ref{signflip}). This ambiguity is taken care of by the requirement that the result satisfies Cutkosky’s rule in eq.(\ref{cutkorule}) which is achieved by fixing the open constants $c_i$ accordingly.
	
	To do this we first find out the discontinuity of eq.(\ref{signflip}) across the $p_2^2$ channel. The HPL's with weight less than or equal to 3 can be represented in terms of classical and Nielsen polylogarithms and hence we can easily find out the discontinuity. Then we can numerically evaluate the original integral at certain kinematical points in order to determine the unknown constants $c_i$.
	The analytical expression of the RHS of eq.(\ref{cutkorule}) is already evaluated and is given by eq.(\ref{finalcut}). We use the \textbf{Ginsh} command of PolyLogTools to find out the numerical value of RHS at the selected kinematical points. Finally, the constant of the weight 4 term which does not appear in the discontinuity, is 
	calculated by numerically evaluating the Feynman integral at a suitable (integral not divergent) kinematical point (integral not divergent). This is in accordance with the general observations in Sec.7 of \cite{Abreu et al.(2014)} in the event of such a procedure is required. The list of the $H_i(z,\bar{z})'s$ and the corresponding $c_i's$ which actually contribute to our result is given in Table.[\ref{tab:my_label}].

	Thus eq.(\ref{fullfunc}) is the other main result of this paper, and it is on the same footing as eq.(7.16) of \cite{Abreu et al.(2014)}. Thus we have completed the non-planar extension of the planar results of \cite{Abreu et al.(2014)}.

\begin{table}[h]
  
   \begin{center}
 \begin{tabular}{||c | c || c | c ||} 
 \hline
  $H_i(z,\bar{z})$ & $c_i$ & $H_i(z,\bar{z})$ & $c_i$  \\ [0.5ex] 
 \hline\hline
 $H(0,0,1,z)$ & $-8i\pi$ & $H(1,z) H\left(0,0,\bar{z}\right)$ & $16i\pi$ \\
 \hline
  $H\left(0,0,1,\bar{z}\right)$ & $16i\pi$ & $H\left(1,\bar{z}\right)H(0,0,z)$ & $-8i\pi$  \\ [0.5ex] 
 \hline
$H(1,0,0,z)$ & $8i\pi$ & $H\left(0,\bar{z}\right)H(0,1,z)$ & $-4i\pi$ \\
 \hline
 $H\left(1,0,0,\bar{z}\right)$ & $-16i\pi$ & $H(0,z)H\left(0,1,\bar{z}\right)$ & $-4i\pi$ \\
 \hline
 $H(1,0,z)$ & $-4\pi^2$ & $H\left(0,\bar{z}\right)H(1,0,z)$ & $12i\pi$\\
 \hline
 $H\left(1,0,\bar{z}\right)$ & $4\pi^2$ & $H(0,z)H\left(1,0,\bar{z}\right)$ & $-12i\pi$ \\
 \hline
 $H(0,1,z)$ & $4\pi^2$ & $H\left(0,\bar{z}\right)H(1,z)$ & $-4\pi^2$ \\
 \hline
 $H\left(0,1,\bar{z}\right)$ & $-4\pi^2$ & $H(0,z)H\left(0,\bar{z}\right)$ & $-16\pi^2$  \\
 \hline
 $H(0,0,z)$  & $-8\pi^2$ & $H(0,z)H\left(1,\bar{z}\right)$ & $4\pi^2$ \\
 \hline
 $H\left(0,0,\bar{z}\right)$ & $-32\pi^2$ & $H(0,z)$ & $-8i\pi^3$ \\
 \hline
 1 & $4\pi^4$ & $H\left(0,\bar{z}\right)$& $-16i\pi^3$\\
 \hline
\end{tabular}
\end{center}
    \caption{Table of $H_i(z,\bar{z})'s$ and the corresponding $c_i's$}
    \label{tab:my_label}
\end{table}

	%%%%%%%%%%%%%%%%%%%%%%%%%%%%%%%%%%%%%%%%%%%%%%%%%%%%%%%%%%%%%%%%%%%%%%%%%%%%%%%%%%%%%%%
	
	\section{Discussion and Conclusions}
	
 In \cite{Abreu et al.(2014)},  inspired by the work of Duhr \cite{Duhr:2012fh}, Abreu et al. have initiated a new method of calculating Feynman integrals which is based on cut diagrams and Hopf algebras. While the method worked well for planar
        two-loop diagrams, its extension to non-planar diagrams and also to more than two loops did not seem clear. In the present paper, therefore, we have considered a non-planar two-loop diagram to study the new method in this case as well. In particular, we have considered the non-planar ladder diagram of \cite{Ussyukina Davydychev(1993)} which is well studied by other methods and can be conveniently compared to the planar ladder. Following the new method,  we have calculated the cuts in the $p_2^2$ channel (see Figure 1) consisting of a two-propagator cut and and three-propagator cut which result in eq.(\ref{firstcutdiv}) - eq.(\ref{firstcutfin}), and eq.(\ref{secondcutf})-eq.(\ref{pre-factor}), respectively. the total cut results are given by eq.(\ref{totalcutdiv})-eq.(\ref{finalcut}) and the total symbol by eq.(\ref{symbol}).  The kinematics of this non-planar problem are very much similar to the planar case in the sense that for the first cut (two-propagators) , the result was obtained by first evaluating a one-loop diagram and using it in the subsequent calculation. Also, in the latter cut, the result of the cut of a one-loop diagram is used as input for the ensuing calculations.

        One difference between the planar and the non planar cases are that in the latter the cut-propagators for the second cut are not co-planar and yet when the ensuing result is added to that coming from the first cut, the total expression is finite. Further, for the first cut, we needed a two-mass easy box rather than a two-mass triangle that was required for the planar case. A noteworthy difference is that we needed to calculate the cut of a three-mass box diagram up to $\eps^1$ order, in contrast to only $\eps^0$ order for the planar case. Also, there are additional contributions from $\eps^{-1}$, $\eps^1$ and the prefactor terms for the second cut. It may be noted that the need to go to these orders in $\eps$ in the non-planar case could not have been foreseen  until the computation was performed.       
        Compared to the planar case, the non-planar calculations were more complex because, for the second cut, we have to do three integrations after expansion of the hypergeometric ${}_2F_1$ functions via HypExp. Without using PolyLogTools, it would have been a challenging task. In \cite{Abreu et al.(2015)}, a specific technique for treating the cuts of the massive internal lines has been introduced.  We have now adopted the same technique for the massless case and applied it to get the desired results.

        It was further discussed in \cite{Abreu et al.(2014)}, that there is no known algorithm to find out the symbol alphabet, which is $\cA=\{z,\zbar,1-z,1-\zbar\}$ for the non-planar case. Due to the result in \cite{Ussyukina Davydychev(1993)} it was possible to obtain this by inspection. The symbol turns out to be equal to that of the one-loop triangle and the planar two-loop ladder diagram discussed in \cite{Abreu et al.(2014)} owing to the fact that they fall in the same family of functions \cite{{Ussyukina Davydychev(1993)}}.

        In \cite{Abreu et al.(2014)}, it was shown for the planar case, in eq. (7.16) of that paper, that the generalized result or the class of Feynman integrals which can give rise to the same symbol can be obtained.  The search for such a structure in the present case is significantly more complicated and has been carried out and reported in Sec.8.

        The main points of the present work are to show that (i) the Hopf Algebra based method also holds for the non-planar case, thereby proving the Duhr conjecture \cite{Duhr:2012fh} for this case, and (ii) that the full result can be reconstructed from the symbol by deploying several tools. The availability of the result (eq.(30) of \cite{Ussyukina Davydychev(1993)}) has allowed us to verify this calculation\footnote{For the planar case, in \cite{Abreu et al.(2014)}, it has been stated that the result can be obtained by suitable manipulations of HPLs. Using the procedure detailed in this paper for the non-planar case, we have also reproduced their results for the planar case ourselves.}. 

        The present work also demonstrates that with modern codes such as HypExp and PolyLogTools, one can now access more easily kinematic configurations that were largely out of reach before. It is satisfying to see that a deep understanding of the structure of the amplitudes as a sum of HPLs of suitable weight, and by a systematic use of the properties of the PolyLog tools can lead to the full result for complex calculations. It would of course be interesting to consider other non-planar diagrams. It is intriguing to think of transformations that relate the planar and non-planar configurations, which may be possible for the cut versions, thereby providing an answer to many intriguing questions.  The present manuscript is the first step in this direction but far from being the last.     

        It will also be interesting to note the usefulness of the results our work on the study of diagrammatic coaction. The diagrammatic coaction is the generalisation of the coproduct of multiple polylogarithms to complete Feynman graphs. Recently Abreu et al. \cite{diagrammatictwoloop} have defined the diagrammatic coaction starting from one loop and extending it to certain class of (planar) two loop diagrams. The diagrammatic coaction poses a very interesting way of studying Feynman integrals and their cuts, paving the way for an easier calculation, compared to working to higher orders in $\epsilon$ expansion involving multiple polylogarithms. Basically it sums up to all orders in $\epsilon$, and the relation with cuts is upgraded in terms of a representation involving only diagrams. Thus an extension to non-planar diagrams could be a project for the future.
	
	\section*{Acknowledgements}
	DW thanks the Indian Institute
	of Science and the Satish Dhawan Chair Professorship for support during this work. AD thanks Ratan Sarkar, Tanay Pathak, Alam Khan, Souvik Bera, Sumit Banik, and Sudeepan Datta for adding valuable discussions during the course of this work. AD also thanks Prof. Dr. Peter Paule, RISC, for providing the access to Harmonic Sums package for performing some calculations during this work.

	\appendix
	\section{Appendix}
	\subsection{Cancellation of the Divergences}
	Here we show explicitly that the divergences of the obtained total cut cancel out. A similar procedure is done for the individual contributions, which shows each of the contributions is independently divergenceless. As discussed earlier, the verification of the cancellation of the divergences is crucial as it is the basis of all types of contributions we have got in the results. 
	
	First of all, we use the command \textbf{ShuffleRegulate} on the sum of the first and the second cut to separate out the divergences, which gives us the following divergent terms, which are now separated from the finite terms.
	$$
	\text{HPL}(\{1\},1) \left(G\left(\frac{z-1}{z-\bar{z}},1\right) \text{HPL}(\{1\},1)+G\left(\frac{\bar{z}-1}{\bar{z}-z},1\right) \text{HPL}(\{1\},1)-G(1,z) G\left(\frac{z-1}{z-\bar{z}},1\right)-\right.
	$$
	$$
	G\left(1,\bar{z}\right) G\left(\frac{z-1}{z-\bar{z}},1\right)+G(1,z) G\left(\frac{z}{z-\bar{z}},1\right)+G\left(1,\bar{z}\right) G\left(\frac{(z-1) \bar{z}}{z-\bar{z}},1\right)-G(1,z) G\left(\frac{z-z \bar{z}}{z-\bar{z}},1\right)
	$$
	$$
	-G\left(1,\bar{z}\right) G\left(-\frac{\bar{z}}{z-\bar{z}},1\right)+G(0,z) \left(G\left(\frac{z-1}{z-\bar{z}},1\right)+G\left(\frac{\bar{z}-1}{\bar{z}-z},1\right)\right)+G\left(\frac{z-1}{z-\bar{z}},0,1\right)+
	$$
	$$
	G\left(0,\bar{z}\right) \left(G\left(\frac{z-1}{z-\bar{z}},1\right)+G\left(\frac{\bar{z}-1}{\bar{z}-z},1\right)\right)+G\left(\frac{z-1}{z-\bar{z}},1,1\right)+G\left(\frac{z}{z-\bar{z}},\frac{z-1}{z-\bar{z}},1\right)+G\left(\frac{\bar{z}-1}{\bar{z}-z},0,1\right)
	$$
	\begin{equation}
		\left.
		-G\left(\frac{(z-1) \bar{z}}{z-\bar{z}},\frac{z-1}{z-\bar{z}},1\right)+G\left(\frac{z-z \bar{z}}{z-\bar{z}},\frac{\bar{z}-1}{\bar{z}-z},1\right)-
		G\left(-\frac{\bar{z}}{z-\bar{z}},\frac{\bar{z}-1}{\bar{z}-z},1\right)+G\left(\frac{\bar{z}-1}{\bar{z}-z},1,1\right)\right)
	\end{equation}
	
	Here the term  $\text{HPL}(\{1\},1)$ is actually divergent, the rest of the expressions are finite. So we just need to check that whether these finite terms are canceling.
	
	The first two terms are nothing but simple logarithms multiplied with $\text{HPL}(\{1\},1)$, and it is very trivial to show that they cancel each other.
	
	For the rest of the terms, we use the command \textbf{Ginsh} of PolyLogTools to check the result for different values of $z$ and $\zbar$. It turns out that the result is always zero irrespective of what values of $z$ and $\zbar$ we take, which shows that these finite terms cancel and hence also the divergences. We can also rewrite them in terms of normal polylogarithms using the definitions of the multiple polylogarithms and check they add up to zero using dilogarithm identities.
	
	\subsection{Realization of Symbol of the Feynman Integral}
	
	As discussed earlier after the total cut for a particular channel is obtained we use the first-entry  condition and the integrability condition repeatedly in order to get the symbol. By analysing the final form of the symbol it turns out that atleast for the studied examples the full symbol is actually given by the following framework :
	\begin{equation}
		\text{Symbol}(F) = \Delta_{1,...,1}(F) =  \sum_i p_i^2 \otimes \Cut_{p_i^2} F ,\hspace{1cm}i\neq1
	\end{equation}
	This means that using the first-entry condition, the first entries of all the terms will always be a particular channel, and the next entry will always be the cut of that particular channel. This is confirmed by the fact that for our example, even if we have started with the other $p_3^2$ channel, we will end with the same symbol. Though we do not have a rigorous proof of this, it is true for other examples studied in \cite{Abreu et al.(2014)}.
	
	This also suggests that using the cut for a particular channel, the first-entry condition, and the integrability condition, we can find out the cuts for the other channels. Also, if we know the cuts for all the channels except for anyone channel, we can find out the original symbol without using the first-entry condition and the integrability condition explicitly.

\end{document}